\documentclass[preprint,12pt]{elsarticle}

\usepackage{amssymb}
\usepackage{epsfig}        
\usepackage{graphicx}	
\usepackage{xcolor}
\usepackage{hyperref}
\usepackage{amsmath}

\newcommand{\bd}[1]{ \mbox{\boldmath $#1$}}

\journal{Nuclear Physics A}

\begin{document}

\begin{frontmatter}

\title{Application of the cranking method to the semimicroscopic algebraic cluster model and nuclear molecules}

\author{D.S. Lohr-Robles$^{1}$, G.E. Morales-Hern\'andez$^{1}$, E. L\'opez-Moreno$^{2}$, P.O. Hess$^{1,3}$}

\address{
{\small\it
$^1$ Instituto de Ciencias Nucleares, Universidad Nacional Aut\'onoma de M\'exico,}\\
{\small\it A.P. 70-543, 04510 Mexico-City, Mexico}\\
{\small\it $^2$ Facultad de Ciencias,  Universidad Nacional Aut\'onoma de M\'exico,} \\
{\small\it 04510 Mexico-City, Mexico} \\
{\small\it $^3$ Frankfurt Institute for Advanced Studies, J. W. von Goethe University, Hessen, Germany}
}

\begin{abstract}
A particular quantum phase transition (QPT) is studied at excited energies of light nuclei within the \textit{Semimicroscopic Algebraic Cluster Model} (SACM), using a combination of catastrophe theory and a direct minimization of the potential. A distinct change from a compact nucleus to a nuclear molecule is described as a second order QPT occurring at a certain value of angular momentum. This finding is in accordance with experimental observations when nuclear molecules appear. The method has implication in the fission of heavy nuclei.
\end{abstract}


\end{frontmatter}

\section{Introduction}
\label{intro}
The study of phase transitions plays an important role in 
nuclear systems \cite{cejnar2010}, in particular in 
nuclear cluster systems. There are two 
possible ways
to study these
transitions: In ground state phase transitions, 
the parameters of a 
particular Hamiltonian are varied
and possible phase transitions are investigated. 
The spectrum of various nuclei is then adjusted and 
it is determined if a phase transition took place from one nucleus to the next one 
\cite{cejnar2010,lopezmoreno1996b,lopezmoreno1998}. 
The second possibility, which 
will be followed here, is to adjust the low energy spectrum of
a particular nucleus and then start to rotate it, in order to study
a possible phase transition at higher energy. This process
is referred as {\it cranking}, a method which plays an
important role in the study of nuclei at excited energies
\cite{ring}.

Two questions then arise: What are the structural changes of the nuclear system at the phase transition? What are the characteristics of the phase transitions (e.g. its order)? 

Depicting a nucleus as two clusters 
within a cluster model is of specific
advantage. At low energy the nucleus is compact,
often deformed, 
and can be treated within
a cluster model, even though the cluster structure is not explicitly seen, which is called the weak definition of clusterization.
At high energy, however, the two clusters may show an explicit
separation, which is called the strong definition of clusterization, resulting in a nuclear molecule. An example of this distinction is
${}^{24}\mathrm{Mg}$, which at low energy is a deformed but compact nucleus,
described via a ${}^{12}\mathrm{C}+{}^{12}\mathrm{C}$ cluster 
structure \cite{cseh1993}, but at high energy it 
transforms into a well established
molecular structure (see, for example,
\cite{cseh1993} and references therein).

The question is now if this transition from a compact to a molecular structure can be retrieved using the cranking
mechanism. This is the objective of this contribution:
A unified geometric description of phases in the low and high 
energy regimes.
We will show that the onset of the formation of nuclear
molecules 
can be described qualitatively within the {\it Semimicroscopic
Algebraic Cluster Model} (SACM). However, though the
onset of the phase transition 
will be 
reproduced, the behavior
of the excited, molecular states present a problem: 
The stable rotation axis is around an axis with the largest
moment of inertia, which leads to the lowest lying energy states,
an axis which cannot be reached by the cranking formalism used
here. The rotation axis is still the x-axis, while the
correct one passes through the touching point of the two
clusters and is orthogonal to the axis connecting the center
of both nuclei, inclined to the old z-axis by an angle 
$\theta$. The obtained band by the cranking formalism
will produce a band with a too low moment of
inertia and too high lying states. We will comment on this
later on.
We restrict 
ourselves to
light cluster systems, where a wealth of experimental information on nuclear molecular systems is available and also where the SACM is well established.
The results, however, are of importance for some heavy
nuclei, when they undergo a fission process. 
In \cite{chavez2021} 
the SACM was extended 
to heavy nuclei and the 
results and the methods developed here will help to study the fission process in future.
It remains to be mentioned that in \cite{NM1,NM2} a geometric model
of nuclear molecules was constructed, including the determination
of its spectrum. It is 
applicable also to light nuclear systems, which is of
importance in order to make a comparison with the geometrical mapping of
the SACM \cite{hess1996} and its spectrum.

This contribution is organized as follows: 
In Section
\ref{summarysacm}
the SACM and the concept of the geometrical
mapping are briefly reviewed. In Section 
\ref{cranking} the cranking mechanism is introduced
into the SACM and the angular momentum will be related to
the {\it cranking parameter} $\Omega$. First, the case $\Omega = 0$ is discussed, which is mainly the result of 
\cite{lohrrobles2019,lohrrobles2021}. Then, small
$\Omega$ values are considered in order to understand 
the evolution of the structure of the cluster system
when $\Omega$ is increased. In Section 
\ref{finiteomega} three systems
will be studied (${}^{12}\mathrm{C}+^{12}\mathrm{C} \to {}^{24}\mathrm{Mg}$,
${}^{12}\mathrm{C}+^{16}\mathrm{O} \to {}^{28}\mathrm{Si}$ and 
${}^{16}\mathrm{O}+^{16}\mathrm{O} \to {}^{32}\mathrm{S}$) which represent
different types of cluster systems and where experimental
data on nuclear molecules are available
\cite{cindro1981,abbondanno1991}. At large
angular momentum the phase transitions found with the cranking formalism are identified with 
the
transition to nuclear molecules.
Finally, in Section \ref{conclusions} conclusions are drawn.

\section{Brief summary of the SACM and the semiclassical potential}
\label{summarysacm}
In the SACM \cite{cseh1992,cseh1994} the internal structure of the clusters is described by the $\mathrm{SU}(3)$ shell 
model \cite{elliott1958a,elliott1958b} and the relative motion 
by the $\mathrm{U}(4)$ vibron model \cite{vibron1,vibron2,vibron3}. For the relative motion the dynamics is determined by $\pi$ bosons with $\ell =1$ angular momentum, and a cut-off is introduced 
through the addition of $\sigma$ bosons with $\ell =0$ angular momentum. The total number of bosons is
required to be always constant $N=n_{\pi}+n_{\sigma}$. The generators of the $\mathrm{U}(4)$ group are the operators: ${\bd \pi}_m^{\dagger}{\bd \pi}^{m'}$, ${\bd \sigma}^{\dagger}{\bd \pi}^m$, ${\bd \pi}_m^{\dagger}{\bd \sigma}$, ${\bd \sigma}^{\dagger}{\bd \sigma}$, which preserve the number of bosons, 
where ${\bd \pi}_m^{\dagger}$ (${\bd \sigma}^{\dagger}$) and ${\bd \pi}^m$ (${\bd \sigma}$) are the creation and annihilation operators of the $\pi$ ($\sigma$) bosons, respectively.

The dynamics of the complete nucleus has a group structure 
\begin{eqnarray}
&
\mathrm{SU}_{C_1}(3)\otimes 
\mathrm{SU}_{C_2}(3)\otimes\mathrm{U}_R(4)  \supset
&
\nonumber \\
&
{\rm SU}_C(3) \otimes {\rm SU}_R(3) 
\supset {\rm SU}(3) \supset {\rm SO}(3)
&
~~~,
\label{dyn-group}
\end{eqnarray}
where the subscript $R$ refers to the relative part and $C_k$ to the $k$-th cluster. 
The $\mathrm{SU}_C(3)$ corresponds to the 
combined cluster
structure and $\mathrm{SU}(3)$ to the total oscillator {\it irreducible
representation} (irrep) of the cluster
system, while $\mathrm{SO}(3)$ is the angular momentum group.

For the relative motion two main dynamical symmetry
group chains exist which contain the $\mathrm{SO}_R(3)$ group for the rotational degree of freedom:
\begin{align}\label{1.1}
&\mathrm{U}_R(4)\supset\mathrm{SU}_R(3)\supset \mathrm{SO}_R(3) \cr
&\mathrm{U}_R(4)\supset\mathrm{SO}_R(4)\supset \mathrm{SO}_R(3)
~~~.
\end{align}

The phenomenological Hamiltonian is a linear combination
of the 
Casimir operators of $\mathrm{SU}(3)$, $\mathrm{U}_R(4)$ 
and $\mathrm{SO}(4)$.
In the present study we will omit the $\mathrm{SO}_R(4)$ dynamical symmetry and consider a more simple 
pure $\mathrm{SU}(3)$ Hamiltonian:
\begin{equation}\label{1}
{\bd H} = \hbar \omega {\bd n}_{\pi} +(a - b\Delta {\bd n}_{\pi}){\bd C}_2(\lambda,\mu) +\xi{\bd L}^2 + t_1 {\bd K}^2
\end{equation}
with $\Delta {\bd n}_{\pi}= {\bd n}_{\pi}-n_0$. The Hamiltonian depends on four parameters $\{a,b,\xi,t_1\}$ given in $\mathrm{MeV}$ units and $\hbar \omega = 45 A^{-1/3}-25A^{-2/3}$ 
\cite{blomqvist1968}, where $A$ is the number of nucleons of the total nucleus. The ${\bd C}_2(\lambda,\mu)$ is the second
order Casimir operator of $\mathrm{SU}(3)$, ${\bd L}^2$ is the square
of the angular momentum and ${\bd K}$ gives the projection
of the angular momentum onto the $z$-axis.

The space of the model is constructed by taking the direct product of the $\mathrm{SU}(3)$ irrep of the relative motion and the individual clusters ground state irreps:
\begin{equation}\label{1.2}
(\lambda_1,\mu_1)\otimes(\lambda_2,\mu_2)\otimes (n_{\pi},0) = \sum m_{\lambda,\mu} (\lambda,\mu),
\end{equation}
where $m_{\lambda,\mu}$ is the multiplicity of the particular irrep, and $n_{\pi}$ is the number of the relative 
oscillation quanta. This number is bounded from below $n_{\pi}\geq n_0$ by the Wildermuth condition \cite{wildermuth}, which is a necessary condition to satisfy the Pauli exclusion principle. The sum of irreps obtained from the product is then compared to the $\mathrm{SU}(3)$ shell model irreps of the total nucleus and only the overlapping irreps are kept. This ensures that the Pauli exclusion principle is taken into account. 
The eigenfunctions of the Hamiltonian are labelled by the quantum numbers of the $\mathrm{SU}(3)$ dynamical symmetry chain 
\begin{eqnarray}
|\left[(\lambda_1,\mu_1)(\lambda_2,\mu_2)\right]
(\lambda_C,\mu_C)(n_\pi,0);(\lambda,\mu), L, M, K \rangle
~~~.
\label{state}
\end{eqnarray}

The total number of relative oscillation quanta is $n_{\pi}=n_0,n_0+1,\ldots,n_0+N$, where $N$ ideally grows to infinity. However, in practice, when diagonalizing the Hamiltonian, in order to fit the parameters to experimental data, a relative small value of $N$ is sufficient to obtain meaningful results. In all three examples considered in the present paper, when fitting the parameters, we used $N=4$, i.e. four additional excitation quanta 
(shell excitations) are added to the complete nucleus.

To study quantum phase transitions in the SACM we construct a semiclassical potential by calculating the expectation value of the Hamiltonian in (\ref{1}) and use as a test function the coherent state \cite{hess1996}:
\begin{align}\label{m.1}
|\alpha \rangle &= \mathcal{N}_{N,n_0}(\boldsymbol{\alpha}^{*}\cdot\boldsymbol{\pi}^{\dagger})^{n_0}[\boldsymbol{\sigma}^{\dagger}+(\boldsymbol{\alpha}^{*}\cdot\boldsymbol{\pi}^{\dagger})]^{N}|0\rangle \cr
&=\left. \frac{N!}{(N+n_0)!}\mathcal{N}_{N,n_0}\frac{d^{n_0}}{d \gamma^{n_0}}[\boldsymbol{\sigma}^{\dagger}+\gamma (\boldsymbol{\alpha}^{*}\cdot\boldsymbol{\pi}^{\dagger})]^{N+n_0}|0\rangle \right|_{\gamma=1}
\end{align}
with the normalization constant
\begin{equation}\label{m.2}
\mathcal{N}^{-2}_{N,n_0}= \left.\frac{(N!)^2}{(N+n_0)!}\frac{d^{n_0}}{d \gamma^{n_0}_1}\frac{d^{n_0}}{d \gamma^{n_0}_2}[1+\gamma_1\gamma_2(\boldsymbol{\alpha}^{*}\cdot\boldsymbol{\alpha})]^{N+n_0}\right|_{\gamma_1=\gamma_2=1} .
\end{equation}
For the {\it semiclassical analysis} the number of total 
relative oscillation quanta $N+n_0$ should also grow to 
infinity. However, while for adjusting the spectrum the
maximal number of $N$ was 4 (sufficient to obtain a good
fit for the parameters), for the study of a phase transition
at higher energy, this is not sufficient, i.e., the $N$ has
to be much larger. In the subsequent sections we will consider 
$N$ finite and equal to a large number, because 
for $N$ sufficiently large the separatrices in parameter space 
begin to maintain their shape and convergence is reached.

The parameters ${\bd \alpha}$ of the coherent state are
treated as arbitrary complex parameters. We will consider the following parametrization 
\cite{lohrrobles2019,moraleshernandez2012,lopezmoreno2016,
morales2012,morales2015}

\begin{align}\label{2}
\alpha_{\pm 1} &= \frac{\alpha}{\sqrt{2}}e^{\pm i\phi}\sin \theta \cr
\alpha_{0} &= \alpha \cos\theta ,
\end{align}
with the variable domains: $\alpha \in[0,\infty)$, $\theta\in[0,\pi]$, and $\phi\in[0,2\pi)$. The variable $\alpha$ is related to the distance between the clusters \cite{hess1996}, while the angular variables $(\theta,\phi)$ represent the orientation of the intercluster axis with the $z$-axis. The $\phi$ is the azimuthal angle, describing rotations
around the $z$-axis, and $\theta$ is the polar angle,
describing the relative orientation of the axis, connecting
both nuclei, to the z-axis.

Before proceeding, we have to say some words on the meaning
of $\alpha$. The semiclassical potential is to be minimized in the coherent state variables. When the global minimum is located at $\alpha = 0$ it is often called the
``spherical'' limit, even though 
within the SACM the nucleus is not spherical,
it is still deformed and its deformation value is
determined by the $(\lambda , \mu )$
irrep, i.e. $\alpha =0$ corresponds to the ground state deformation
of the nucleus
regime. Also, the distance between the clusters is 
{\it finite} and determined by the minimal number
of quanta \cite{hess1996} due to the Wildermuth condition.
An increasing $\alpha$ only increases that distance.
Thus, the interpretation of the limit $\alpha \rightarrow 0$
is very different to the interpretation in the nuclear
vibron model, here it does not correspond to the vibrational 
limit. When the global minimum is located at some $\alpha>0$ the total deformation
increases, so that in the phase transition the two-cluster system is
strongly deformed and corresponds
to a molecular state, with a well
defined  separation of the two
clusters.

The semiclassical potential 
$V=\langle \alpha | {\bd H} | \alpha \rangle$ 
results in a function of two variables $(\alpha,\theta)$ 
and is independent of $\phi$
\cite{lohrrobles2019}:
\begin{align}\label{3}
V(\alpha,\theta;a,b,\xi) &= V_0 + \Big(A_0 + A_1 (1+3\cos 2\theta)\Big) \alpha^2 \frac{F_{11}(\alpha)}{F_{00}(\alpha)} \cr
&\quad +\Big(B_0 + B_1 (1+3\cos 2\theta)+ \xi\sin ^2 2\theta\Big) \alpha^4 \frac{F_{22}(\alpha)}{F_{00}(\alpha)} - b  \alpha^6 \frac{F_{33}(\alpha)}{F_{00}(\alpha)},
\end{align}
where $V_0$ and the parameters $\{A_i,B_i\}$ are given by:
\begin{align}\label{4}
V_0&=(a+bn_0)\langle {\bd C}_2(\lambda_C,\mu_C)\rangle + \xi \langle {\bd L}_C^2 \rangle +t_1 \langle {\bd K}^2\rangle\cr
A_0&= \hbar \omega + 4(a+b(n_0-1))-b 
\langle {\bd C}_2(\lambda_C,\mu_C)\rangle + 2\xi \cr
A_1&= \frac{1}{4}(a+b(n_0-1))\langle Q_{C,0}^a \rangle \cr
B_0&= a+b(n_0-6)\cr
B_1&=-\frac{b}{4}\langle Q_{C,0}^a \rangle ,
\end{align}
and the $F_{pq}(\alpha)$ functions defined as 
\cite{yepezmartinez2012a}:
\begin{align}\label{5}
F_{pq}(\alpha^2)&= \frac{(N!)^2}{(N+n_0-\max(p,q))!}\\
&\quad\times\sum_{k=\max(n_0-p,n_0-q)}^{N+n_0-\max(p,q)}\left(
\begin{array}{c}
N+n_0-\max(p,q) \\
k
\end{array}
\right)
\frac{(k+p)!}{(k+p-n_0)!}\frac{(k+q)!}{(k+q-n_0)!}\alpha^{2k} .\nonumber
\end{align}
The deformation of the two cluster
system is taken into account in the expectation value of the $m=0$ component of the cluster quadrupole operator \cite{hess1996}:
\begin{equation}\label{6}
\langle (\lambda_C,\mu_C) | Q_{C,0}^a| (\lambda_C,\mu_C)\rangle = \sqrt{\frac{5}{\pi}} \left(n_C + \frac{3}{2}(A_C -1)\right) \beta_C .
\end{equation}
where $A_C=A=A_1+A_2$, $n_C$ is the total number 
of quanta, $A_k$ is the number of nucleons
within the $k$-th cluster, and $\beta_C$ is the deformation parameter of the combined cluster system. It is assumed that the relative
orientation of the two clusters is maintained.
The irrep $(\lambda_C,\mu_C)$ represents an intermediate irrep obtained by the direct product of the individual clusters irreps: $(\lambda_1,\mu_1)\otimes (\lambda_2,\mu_2)$. The $(\lambda_C, \mu_C)$ is the result of the
coupling of the two clusters, where a smaller irrep 
corresponds to a more compact configuration. For example,
the irrep $(\lambda_1+\lambda_2,\mu_1+\mu_2)$ is a 
linear, mostly deformed configuration. In the
${}^{12}\mathrm{C}+{}^{12}\mathrm{C}$ this is the $(0,8)$ irrep, while the
most compact irrep corresponds to $(4,0)$, which is the smallest
irrep appearing in the product $(0,4) \otimes (0,4)$.
The $\beta_C$ value depends on $(\lambda_C,\mu_C)$ respectively given in \cite{rowe,castanos1988} as:
\begin{eqnarray}\label{betac} 
\beta_C^2 & = & \frac{16\pi}{5 N_0^2}\left( \lambda_C^2 + \lambda_C \mu_C
+\mu_C^2 \right)
~~~. \cr
\beta_C^2 &=& \frac{4\pi}{5 r_0^4 A^{8/3}}\left( \lambda_C^2 + \lambda_C \mu_C
+\mu_C^2  + 3\lambda_C + 3\mu_C + 3 \right),
\end{eqnarray}
where $N_0\approx 0.9 A^{4/3}$ and $r_0\approx 0.87$. Here we will use the second relation given in \cite{castanos1988},
considering that both formulas give deformation values in the same range.

\begin{figure}[ht]
\begin{center}
\includegraphics[scale=0.8]{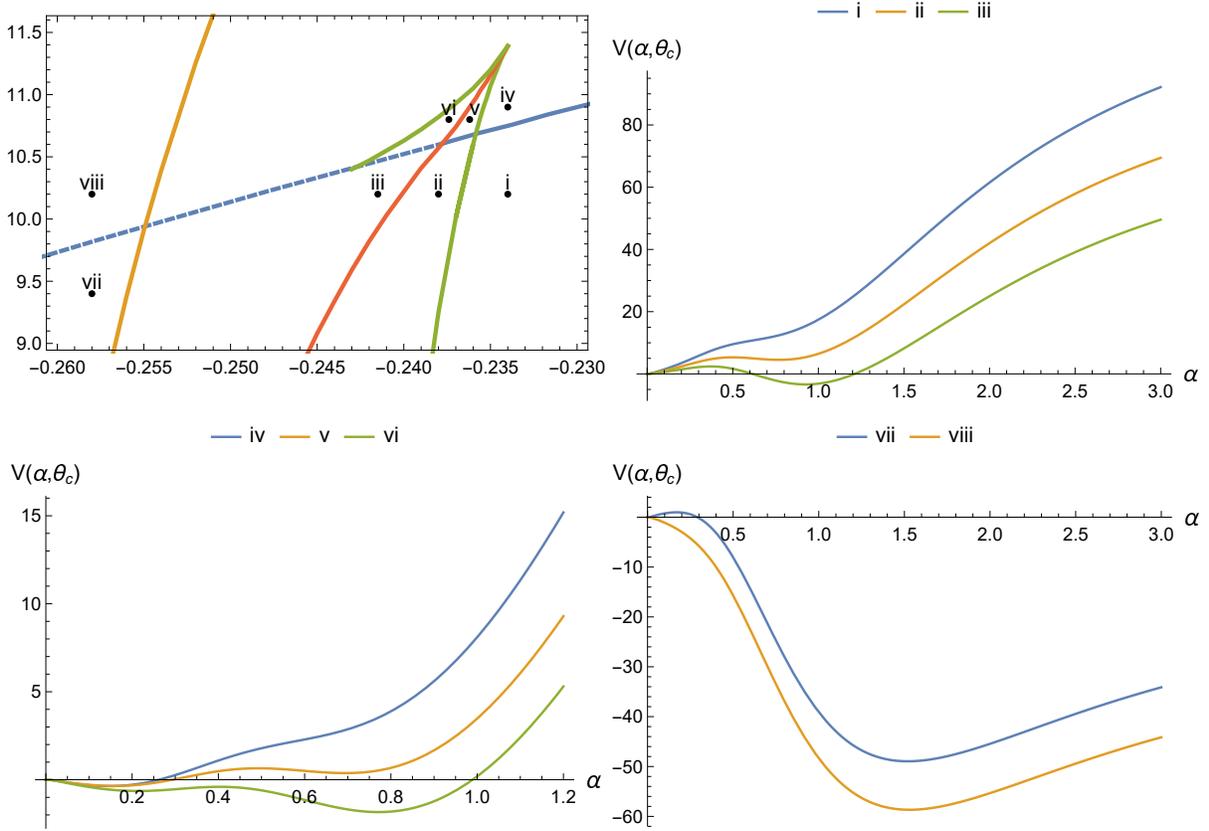}
\end{center}
\caption{Parameter space $(a,\Omega)$ for fixed values $b=-0.002$ and $\xi=0.2$ for the system: ${}^{12}\mathrm{C}+{}^{12}\mathrm{C}$. The second order transition separatrix is depicted as a blue line, with the continuous line representing where the second order QPT takes place. The bifurcation set and Maxwell set are depicted as a green and red line, respectively. The stability separatrix is depicted as a yellow line. 
Potentials are plotted to show the main characteristics of the different regions, and they correspond to the points depicted in the upper left part of the figure.
}
\label{fig.2.1}
\end{figure}

\section{The cranking method and quantum phase transitions}
\label{cranking}
Initially, we consider the cluster system to be 
connected along the $z$-axis. In the cranking method the cluster system is excited by applying a rotation around the $x$-axis with an $\Omega$ angular frequency 
given in $\mathrm{MeV}/\hbar$ units.
This is done by 
extending the Hamiltonian to
\begin{equation}\label{2.1}
{\bd H}_{cranking}={\bd H}-\Omega {\bd L}_x
\end{equation}
where ${\bd H}$ is the Hamiltonian defined in (\ref{1}) and ${\bd L}_x$ is the $x$ component of the angular momentum given by
\begin{equation}\label{2.2}
{\bd L}_x = \frac{1}{\sqrt{2}} \left( {\bd L}_{-1} - {\bd L}_{+1} \right) =  [{\bd \pi}^{\dagger}\otimes {\bd \pi}]_{-1}^{1} - [{\bd \pi}^{\dagger}\otimes {\bd \pi}]_{+1}^{1}.
\end{equation}

The expectation value of ${\bd L}_x$ in the coherent state basis results in \cite{morales2012,morales2015}:

\begin{equation}\label{2.3}
\langle \alpha| {\bd L}_x | \alpha \rangle = \sin 2\theta \cos \phi \alpha^2 \frac{F_{11}(\alpha)}{F_{00}(\alpha)}
~~~,
\end{equation}
which depends on the three variables $(\alpha,\theta,\phi)$. This expectation value is related to the eigenvalue $L$ 
through $\langle \alpha| {\bd L}_x | \alpha \rangle = \sqrt{L(L+1)}$, 
or for large $L$ we can 
approximate it by $\langle \alpha| {\bd L}_x | \alpha \rangle\approx L$, which will be used in the rest of the paper.

It is important to know the limiting cases of (\ref{2.3}) when $\alpha$ goes to zero and when it goes to infinity:
\begin{align}\label{2.4}
\lim_{\alpha\to 0} \alpha^2 \frac{F_{11}(\alpha)}{F_{00}(\alpha)} &=  n_0 \cr
\lim_{\alpha\to \infty} \alpha^2 \frac{F_{11}(\alpha)}{F_{00}(\alpha)} &=  N+n_0 
~~~.
\end{align}
Note that it is bounded and can go from $n_0$ to $N+n_0$, depending on the values taken by $\theta$ and $\phi$.

Taking into account (\ref{2.3}) and (\ref{2.1}), the semiclassical potential to be studied in the cranking formalism is now
\begin{align}\label{2.5}
V(\alpha,\theta,\phi;a,b,\xi,\Omega) &= V_0 + \Big(A_0 + A_1 (1+3\cos 2\theta)-\Omega \sin 2\theta \cos\phi \Big) \alpha^2 \frac{F_{11}(\alpha)}{F_{00}(\alpha)} \cr
&\quad +\Big(B_0 + B_1 (1+3\cos 2\theta)+ \xi\sin ^2 2\theta\Big) \alpha^4 \frac{F_{22}(\alpha)}{F_{00}(\alpha)} - b  \alpha^6 \frac{F_{33}(\alpha)}{F_{00}(\alpha)},
\end{align}
and the procedure consists in minimizing it in the coherent state variables for some values of the parameters $\{a,b,\xi,\Omega\}$. The critical points of (\ref{2.5}) are those that satisfy
\begin{equation}\label{2.6}
\left.\nabla V \right|_{(\alpha_c,\theta_c,\phi_c)}=0
~~~,
\end{equation}
where each component is simultaneously zero for the same particular values of the parameters. The critical point for the variable $\phi$ is readily obtained as $\phi_c=0$. The $\theta$ component of (\ref{2.6}) is given by the following partial derivative:
\begin{equation}\label{2.7}
\left.\frac{\partial V}{\partial \theta}\right|_{\theta_c} = -2 \left[ \left(3A_1\sin 2\theta_c + \Omega \cos 2\theta_c \right)\alpha^2 \frac{F_{11}(\alpha)}{F_{00}(\alpha)} +(3B_1 \sin 2\theta_c - 2\xi \sin 2\theta_c \cos 2\theta_c)\alpha^4 \frac{F_{22}(\alpha)}{F_{00}(\alpha)}\right] = 0.
\end{equation}
In the following subsections we will solve Eq. (\ref{2.7}) for the cases $\Omega =0$ and $\Omega \ll 1$, while the case $\Omega \neq 0$ is solved numerically in the following section where the results are presented. From the study of the $\alpha$ component of (\ref{2.6}) we find that $\alpha_c=0$ is always a critical point, independent of the parameters, and it is called the {\it fundamental root}.

The problem consists in relating the change in the values of the parameters $\{a,b,\xi,\Omega\}$ to the pair of critical points $(\alpha_c,\theta_c)$, with the ultimate goal of constructing separatrices in the parameter space which divide it in regions where the qualitative structure of the potential remains the same. We will see that in the case $\Omega=0$ this task can be done using the methods of catastrophe theory 
\cite{gilmore} in an elegant way. However, in the case $\Omega\neq 0$ we will directly minimize the potential in the $(\alpha,\theta)$ variables for particular sets of the parameters until the conditions for the different separatrices are met. There are four separatrices in the parameter space which are of interest: The second order transition separatrix, the stability separatrix, the bifurcation set and the Maxwell set. In the upper left plot of Fig. \ref{fig.2.1} we show the parameter space $(a,\Omega)$, for fixed values of $b$ and $\xi$, with these four separatrices present, and we also show the potential plots representing the qualitative behaviour of each one of the different regions. 
In what follows we will explain each of the four separatrices and their defining characteristics.

\begin{itemize}

\item {\it Second order transition separatrix}:
We turn our attention to the critical point $\alpha_c=0$ and determine for what values of the parameters does it correspond to a minimum. Expanding the semiclassical potential (\ref{2.5}) in a Taylor series around $\alpha_c=0$, we obtain:
\begin{equation}\label{2.8}
V(\alpha,\theta;a,b,\xi,\Omega) = V_0+T_0(\theta;a,b,\xi,\Omega)+ T_1(\theta;a,b,\xi,\Omega) \alpha^2 + T_2(\theta;a,b,\xi,\Omega) \alpha^4 +\ldots ,
\end{equation}
where the first two coefficients are explicitly given by
\begin{align}\label{2.9}
T_0(\theta;a,b,\xi,\Omega) &= n_0 \Big(A_0+A_1(1+3\cos 2\theta)-\Omega \sin 2\theta \cr
&\quad +(n_0-1)(B_0+B_1(1+3\cos 2\theta)+\xi \sin^2 2\theta) - (n_0-1)(n_0-2)b \Big)\cr 
T_1(\theta;a,b,\xi,\Omega) &= N(n_0+1)\Big(A_0+A_1(1+3\cos 2\theta)-\Omega \sin 2\theta \cr
&\quad +2n_0(B_0+B_1(1+3\cos 2\theta)+\xi \sin^2 2\theta)-3n_0(n_0-1)b\Big).
\end{align}

We define the {\it second order transition separatrix} as the set of parameters for which the $T_1$ coefficient of the Taylor series vanishes, for a critical value $\theta_c$. This separatrix determines when the spherical minimum at $\alpha_c=0$ disappears and becomes a deformed minimum at 
$\alpha_c>0$, i.e., 
when $T_1$ becomes negative. This point represents the value of the parameters where a second order quantum phase transition occurs. This separatrix has also been constructed when applying the cranking method to the interacting boson model (IBM) \cite{cejnar2002,cejnar2003,cejnar2004}. In the IBM the different shapes of nuclei are described by a Hamiltonian of the various dynamical symmetries within the model. The Hamiltonian considered depends on two control parameters and three dynamical symmetry limits are obtained for certain values of the parameters. The dynamical symmetries are: $\mathrm{U}(5)$, $\mathrm{SU}(3)$ and $\mathrm{SO}(6)$, which correspond to spherical nuclei, deformed axially symmetric nuclei and $\gamma$-unstable nuclei, respectively. 
One is able to construct a parameter space in the 
cranking parameter
$\Omega$,
where a separatrix divides the space into spherical and deformed regions and study the $\mathrm{U}(5)-\mathrm{SU}(3)$ transition as the cranking frequency increases. 
In \cite{cejnar2004} the critical frequencies 
are determined for the spherical-deformed transition for some heavy nuclei ($A\approx 100$) and a comparison 
to experimental results is presented.

\item {\it Stability separatrix}:
The stability separatrix determines if the potential for $N$ finite is stable or unstable, and is defined by the limit of the potential as $\alpha\to\infty$:
\begin{align}\label{2.10}
\lim_{\alpha\to\infty} V \equiv T_{inf} &= (N+n_0)\Big(A_0+A_1(1+3\cos 2\theta)-\Omega \sin 2\theta  \cr
&\quad+ (N+n_0-1)(B_0+B_1(1+3\cos 2\theta)+\xi \sin^2 2\theta)\cr
&\quad-b(N+n_0-1)(N+n_0-2)\Big)
~~~.
\end{align} 
The {\it stability separatrix} is the set of parameters for which the value of $T_{inf}$ is equal to the value of the potential at $\alpha_c=0$, i.e. when the following relation is satisfied:
\begin{equation}\label{2.11}
T_{inf}(\theta_{c,\infty};a,b,\xi,\Omega) - T_0(\theta_{c,0};a,b,\xi,\Omega)-V_0 =0,
\end{equation}
where $\theta_{c,\infty}$ denotes 
the value of the critical point of $\theta$ in the limit $\alpha\to\infty$, while $\theta_{c,0}$ is the value of the critical point at $\alpha_c=0$.

\item {\it Bifurcation and Maxwell set}:
As we mentioned, $\alpha_c=0$ is always a critical point and in the region where it corresponds to a minimum there exist a particular set of parameters for which a new minimum appears located at some $\alpha_c>0$ such that $V(\alpha_c=0,\theta_{c,0})<V(\alpha_c>0,\theta_c)$. This subset of parameter space is known as the {\it bifurcation set}. As the value of the parameters are continued to be varied the value of the new minimum starts approaching the value of the minimum at $\alpha_c=0$ until, for a particular set of parameters, it happens that $V(\alpha_c=0,\theta_{c,0})=V(\alpha_c>0,\theta_c)$, i.e. both minima have the same value. This subset in parameter space is known as the {\it Maxwell set}, and it represents 
the value of the parameters where a first order quantum phase transition occurs.

\end{itemize}

The method for finding these separatrices consists in minimizing the potential in the variables $(\alpha,\theta)$ for some values of the parameters $\{a,b,\xi,\Omega\}$. The set of values for which the different conditions, that define the separatrices, are met determines each of the respective separatrices. Then we are able to construct surface separatrices in three-dimensional space, 
e.g. $(a,\xi,\Omega)$ for a fixed value of $b$ or $(a,b,\Omega)$ for a fixed value of $\xi$.

In this contribution we are interested in possible phase
transitions within a particular nucleus which becomes excited 
by applying a rotation.
When studying these particular examples the only free parameter will be the cranking frequency $\Omega$, while the parameters $\{a,b,\xi\}$ in the Hamiltonian (\ref{1}) are going to be fitted to the experimental data at low energy. 
Then, in the two-dimensional space $(a,\Omega)$ with $\{b,\xi\}$ fixed the particular nucleus is represented by a point, and as the cranking frequency $\Omega$ is increased a trajectory in parameter space is traced until one of a the separatrices is crossed and a QPT takes place. We will now turn our attention to the study of the case $\Omega=0$ and also $\Omega\ll 1$ as they are good starting points for this problem.

\begin{figure}[ht]
\begin{center}
\includegraphics[scale=0.8]{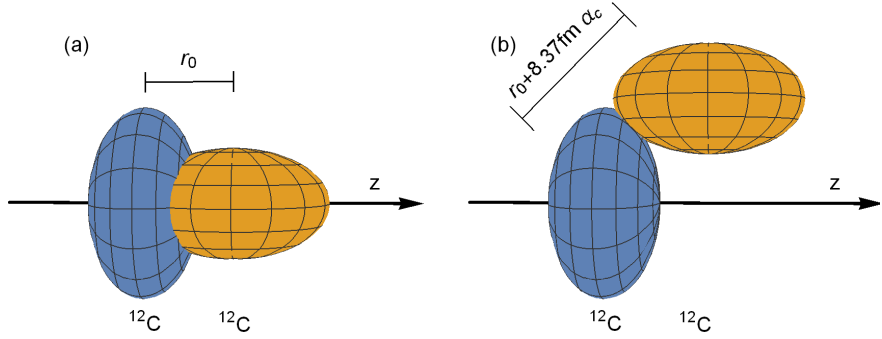}
\end{center}
\caption{Schematic illustration of the system ${}^{12}\mathrm{C}+{}^{12}\mathrm{C}$. In (a) the system is separated by the minimal $r_0 = 2.36 \;\mathrm{fm}$ distance determined by $n_0$ \cite{hess1996}. In (b), after the quantum phase transition, the separation distance of the clusters is larger 
and the new intercluster axis makes an angle $\theta_c \approx \pi/4$ with respect to the $z-$axis}. The two equal semi-axes of the oblate clusters have a value of $2.75\;\mathrm{fm}$, while the radius along the symmetry axis is $1.58\;\mathrm{fm}$. 
\label{fig.4.1.0}
\end{figure}

\subsection{Case: $\Omega =0$.}
\label{omegazero}
This case was studied in \cite{lohrrobles2019,lohrrobles2021}
and for completeness we will summarize some results, important to understand the subsequent study.

From (\ref{2.5}) we obtain the semiclassical potential in this case as:
\begin{align}\label{3.1.1}
V(\alpha,\theta;a,b,\xi,\Omega=0) &= V_0 + \Big(A_0 + A_1 (1+3\cos 2\theta) \Big) \alpha^2 \frac{F_{11}(\alpha)}{F_{00}(\alpha)} \\
&\quad +\Big(B_0 + B_1 (1+3\cos 2\theta)+ \xi\sin ^2 2\theta\Big) \alpha^4 \frac{F_{22}(\alpha)}{F_{00}(\alpha)} - b  \alpha^6 \frac{F_{33}(\alpha)}{F_{00}(\alpha)}, \nonumber
\end{align}
and the condition for critical points of the variable $\theta$ in (\ref{2.7}) becomes:
\begin{equation}\label{3.1.3}
\left.\frac{\partial V}{\partial \theta}\right|_{\theta_c} = -2\sin 2\theta_c \left[ 3A_1  \alpha^2 \frac{F_{11}(\alpha)}{F_{00}(\alpha)} +(3B_1  - 2\xi  \cos 2\theta_c)\alpha^4 \frac{F_{22}(\alpha)}{F_{00}(\alpha)}\right] = 0
~~~.
\end{equation}
We find three different solutions:
\begin{align}\label{3.1.4}
\theta_1 &= 0 \cr
\theta_2 &=\frac{\pi}{2} \cr
\cos 2\theta_3 &= \frac{3A_1}{2\xi} \frac{F_{11}(\alpha)}{\alpha^2 F_{22}(\alpha)} + \frac{3B_1}{2\xi}. 
\end{align}
The first two constant solutions correspond to the cases where the two cluster system is aligned along the $z$-axis or perpendicular to it, respectively, and may be treated together. We renamed them as $\theta_{\pm}$ where $\theta_{+}=0$ and $\theta_{-}=\pi/2$. Direct substitution of $\theta_{\pm}$ in (\ref{3.1.1}) results in the one-dimensional function:
\begin{equation}\label{3.1.5}
V(\alpha,\theta_{\pm};\sigma_i) = V_0   + \sigma_1 \alpha^2 \frac{F_{11}(\alpha)}{F_{00}(\alpha)} + \sigma_2 \alpha^4 \frac{F_{22}(\alpha)}{F_{00}(\alpha)} + \sigma_3 \alpha^6 \frac{F_{33}(\alpha)}{F_{00}(\alpha)}
\end{equation}
and the new parameters $\sigma_i$ are defined as:
\begin{align}\label{3.1.6}
\sigma_1 &= A_0 + A_1 (1\pm 3) = \hbar \omega + 4(a+b(n_0-1))-b \langle {\bd C}_2(\lambda_c,\mu_c)\rangle + 2\xi \cr
&\quad+ \frac{1\pm 3}{4}(a+b(n_0-1))\langle Q_{C,0}^a \rangle \cr
\sigma_2 &= B_0 + B_1 (1\pm 3) = a+b(n_0-6) - b\frac{1\pm 3}{4}\langle Q_{C,0}^a \rangle \cr
\sigma_3 &= -b .
\end{align}

It is convenient to define the $r_i$ parameters following Ref. \cite{lohrrobles2021}:
\begin{align}\label{3.1.7}
r_1 &= -\frac{1}{b}\left(\sigma_1 + 2n_0 \sigma_2 + 3n_0(n_0-1)\sigma_3 \right)\cr
r_2 &= -\frac{1}{b}\left(\sigma_2 + 3n_0 \sigma_3 \right)
~~~,
\end{align}
and subtracting $V_0$ and the first coefficient $T_0$ of the Taylor series in (\ref{2.9}) evaluated at $\theta_{\pm}$, we can rewrite the potential as
\begin{equation}\label{3.1.8}
V(\alpha;r_i)= \frac{-b}{Q_0(\alpha)}\left[ r_1 Q_1 (\alpha) + r_2 Q_2 (\alpha)+  Q_3 (\alpha) \right],
\end{equation}
which vanishes at $\alpha=0$. The polynomials $Q_i(\alpha)$ are defined as
\begin{equation}\label{3.1.9}
Q_i(\alpha) = \sum_{k=i}^N \frac{N!}{(N-k)!} \frac{(n_0+k)!}{n_0!} \frac{\alpha^{2k}}{k!(k-i)!}.
\end{equation}

Using the methods of catastrophe theory \cite{gilmore} described in the Appendix of Ref. \cite{lohrrobles2021}, we can construct the bifurcation and Maxwell separatrices. Along with the stability separatrix they divide the parameter space in regions of similar qualitative behaviour. After the Hamiltonian for each example system is fitted to experimental data and the parameters $(a,b,\xi)$ are determined, we are able to map them to the the $(r_1,r_2)$ parameter space an see in which region they fall. It is important to find out first which of the $\theta$ critical points corresponds to the dominant minimum of the potential surface $V(\alpha,\theta;a,b,\xi,\Omega=0)$ for the particular values of the parameters that result in the fitting. For now we will turn our attention to the $\Omega \neq 0$ case and study first the case for small values of $\Omega$.

\subsection{Case: $\Omega \ll 1$.}
\label{smallomega}
At the end of this subsection we will be able to 
obtain an analytical formula for the moment of inertia for small values of $\Omega$. For the perturbation analysis it is convenient to introduce the following change of variable:
\begin{align}\label{3.2.1}
\gamma &= \cos 2\theta \cr
\frac{\partial}{\partial \theta} &=\pm \sqrt{1-\gamma^2} \frac{\partial}{\partial \gamma} .
\end{align}

With this change of variable the semiclassical potential in (\ref{2.5}) can be written as:
\begin{align}\label{3.2.2}
V(\alpha,\gamma;a,b,\xi,\Omega) &= V_0 + \Big(A_0 + A_1 +3A_1 \gamma \mp\Omega \sqrt{1-\gamma^2} \Big) \alpha^2 \frac{F_{11}(\alpha)}{F_{00}(\alpha)} \cr
&\quad +\Big(B_0 + B_1 +3 B_1\gamma+ \xi (1-\gamma^2)\Big) \alpha^4 \frac{F_{22}(\alpha)}{F_{00}(\alpha)} - b  \alpha^6 \frac{F_{33}(\alpha)}{F_{00}(\alpha)},
\end{align}
and the condition for critical points of the $\theta$ variable in (\ref{2.7}) becomes
\begin{equation}\label{3.2.3}
\left.\pm\sqrt{1-\gamma^2} \frac{\partial V}{\partial \gamma} \right|_{\gamma_c} = \pm\sqrt{1-\gamma_c^2}\left[ \left(3 A_1 \pm \Omega \frac{\gamma_c}{\sqrt{1-\gamma_c^2}} \right)\alpha^2 F_{11}(\alpha) + \left(3B_1 - 2\xi \gamma_c \right) \alpha^4 F_{22}(\alpha) \right] =0 .
\end{equation}
Again, for the case $\Omega=0$ we obtain three solutions:
\begin{align}\label{3.2.4}
\gamma_1 &= 1 \cr
\gamma_2 &= -1 \cr
\gamma_3 &= \frac{3A_1}{2\xi} \frac{F_{11}(\alpha)}{\alpha^2 F_{22}(\alpha)} + \frac{3B_1}{2\xi}
\end{align}
which respectively map to the solutions in (\ref{3.1.4}).

We treat $\Omega$ as a perturbation and suppose that the solutions of Eq. (\ref{3.2.3}) can be written as a power series in $\Omega$ as:
\begin{equation}\label{3.2.5}
\gamma_c = \sum_{k=0}^{\infty} c_k \Omega^k .
\end{equation}
After some algebraic manipulation of Eq. (\ref{3.2.3}), in order to put it in a suitable form, we directly substitute the solution (\ref{3.2.5}) to determine the coefficients $c_k$. Up to second order in $\Omega$ we obtain the solutions:
\begin{align}
\gamma_1 &= 1 - \frac{\Omega^2}{2(3B_1 -2\xi)^2} \frac{\left(\frac{ F_{11}(\alpha)}{\alpha^2 F_{22}(\alpha)}\right)^2}{\left(1+ \frac{3A_1}{3B_1 -2\xi} \frac{F_{11}(\alpha)}{\alpha^2 F_{22}(\alpha)}\right)^2} + \mathcal{O}(\Omega^3) \label{3.2.6a}\\
\gamma_2 &= -1 + \frac{\Omega^2}{2(3B_1 +2\xi)^2} \frac{\left(\frac{ F_{11}(\alpha)}{\alpha^2 F_{22}(\alpha)}\right)^2}{\left(1+ \frac{3A_1}{3B_1 +2\xi} \frac{F_{11}(\alpha)}{\alpha^2 F_{22}(\alpha)}\right)^2} + \mathcal{O}(\Omega^3) \label{3.2.6b} \\
\gamma_3 &= \frac{1}{2\xi} \left(3A_1\frac{F_{11}(\alpha)}{\alpha^2 F_{22}(\alpha)} + 3B_1 \right) + \frac{\Omega}{4\xi^2} \frac{ \frac{ F_{11}(\alpha)}{\alpha^2 F_{22}(\alpha)}\left(3A_1\frac{F_{11}(\alpha)}{\alpha^2 F_{22}(\alpha)} + 3B_1 \right)}{\sqrt{1- \frac{1}{4\xi^2}\left(3A_1\frac{F_{11}(\alpha)}{\alpha^2 F_{22}(\alpha)} + 3B_1 \right)^2}} \cr
&\quad + \frac{\Omega^2}{8\xi^3}\frac{ \left(\frac{ F_{11}(\alpha)}{\alpha^2 F_{22}(\alpha)}\right)^2\left(3A_1\frac{F_{11}(\alpha)}{\alpha^2 F_{22}(\alpha)} + 3B_1 \right)}{\left(1- \frac{1}{4\xi^2}\left(3A_1\frac{F_{11}(\alpha)}{\alpha^2 F_{22}(\alpha)} + 3B_1 \right)^2\right)^2}+\mathcal{O}(\Omega^3) \label{3.2.6c}.
\end{align}

We recall the expectation value of the $x$ component of angular momentum within the basis of coherent states in (\ref{2.3}). 
Taking into account that $\phi_c=0$ is the critical point of the $\phi$ variable and that 
$\sin 2\theta = \pm \sqrt{1-\gamma^2}$, we obtain
\begin{equation}\label{3.2.8}
\langle \alpha | {\bd L}_x | \alpha \rangle = \sqrt{1-\gamma^2}\alpha^2 \frac{F_{11}(\alpha)}{F_{00}(\alpha)} ,
\end{equation}
where we considered the positive root.

To the lowest order in $\alpha$ and $\Omega$, we substitute $\gamma_1$ in (\ref{3.2.8}) and obtain
\begin{equation}\label{3.2.9}
\langle \alpha | {\bd L}_x | \alpha \rangle = \frac{n_0 \Omega}{3A_1 + (n_0-1)(3B_1 - 2\xi)} \left(1+ \mathcal{O}(\alpha^2) \right) + \mathcal{O}(\Omega^2).
\end{equation}
Similarly we substitute $\gamma_2$ in (\ref{3.2.8}) and get
\begin{equation}\label{3.2.10}
\langle \alpha | {\bd L}_x | \alpha \rangle = \frac{n_0 \Omega}{3A_1 + (n_0-1)(3B_1 + 2\xi)} \left(1+ \mathcal{O}(\alpha^2) \right) + \mathcal{O}(\Omega^2).
\end{equation}

Using the relation 
$L=I\Omega +\mathcal{O}(\Omega^2)$ \cite{schaaser1986}, and the approximation $\langle \alpha | {\bd L}_x | \alpha \rangle\approx L$ and comparing it to (\ref{3.2.9}) and (\ref{3.2.10}), we identify the moment of inertia for small value of $\Omega$ as
\begin{align}\label{3.2.11}
I_{\pm} &= \frac{n_0 }{3A_1 + (n_0-1)(3B_1 \mp 2\xi)}  + \mathcal{O}(\Omega) = \frac{n_0}{\frac{3}{4}a \langle Q_{C,0}^a \rangle \mp 2(n_0 -1) \xi} + \mathcal{O}(\Omega),
\end{align}
where $I_{+}$ corresponds to the case where $(\alpha_c=0,\theta_c=0)$ is the dominant minimum and $I_{-}$ to the case where $(\alpha_c=0,\theta_c=\pi/2)$ is the dominant minimum. In the last step of (\ref{3.2.11}) we used the explicit values of $A_1$ and $B_1$ from (\ref{4}). We notice that the moment of inertia depends on the deformation of the two cluster system, as explained
further above.

Therefore, if we are dealing with system of spherical clusters, or clusters with a small value of the deformation parameter, the moment of inertia reduces to:
\begin{equation}\label{3.2.13}
I_{-} =\frac{n_0}{n_0-1} \frac{1}{2\xi} + \mathcal{O}(\Omega) \approx \frac{1}{2\xi} 
~~~,
\end{equation}
to the lowest order, which has already been obtained for the SACM in Ref. \cite{yepezmartinez2008}.

\section{Finite $\Omega$ and the emergence of nuclear molecules}
\label{finiteomega}
In this section we will apply the method described in the previous section to three examples: A symmetric system with two deformed clusters ${}^{12}\mathrm{C}+{}^{12}\mathrm{C}\to {}^{24}\mathrm{Mg}$, a non-symmetric system with one deformed and one spherical cluster ${}^{12}\mathrm{C}+{}^{16}\mathrm{O}\to {}^{28}\mathrm{Si}$, and a symmetric system with two spherical clusters ${}^{16}\mathrm{O}+{}^{16}\mathrm{O}\to {}^{32}\mathrm{S}$. However, in the last system the concept of forbiddenness \cite{smirnov1984,yepezmartinez2015} needs to be taken into account resulting in the system no longer being symmetric and one cluster becomes effectively deformed. 

The main aspect of forbiddenness is as follows: As long as the ground state irrep is reached
in the product $(\lambda_1,\mu_2) \otimes (\lambda_2,\mu_2)
\otimes (n_\pi ,0)$, the clusters are not excited (their irreps are fixed), 
because increasing the excitation quanta in
a cluster is equivalent to exciting only the relative motion.
However, when the ground state irrep is not reached, one can start to shift relative excitation quanta to the clusters.
I.e., one by one a quantum
from the relative motion is subtracted and added to the
cluster system, until the ground state irrep 
of the united nucleus can be reached
for the first time. After that, no more quanta
are subtracted from the relative motion, 
because it would lead again to a double counting.
This concept was introduced in \cite{smirnov1984} and helped
to understand how a system can be divided into a cluster
and which combinations are preferred in certain reactions.

In the three examples proposed a phenomenon 
can be studied, where for certain energy regions the nuclei are identified as nuclear molecules, where the clusters are notably separated.

\begin{figure}[ht]
\begin{center}
\includegraphics[scale=0.8]{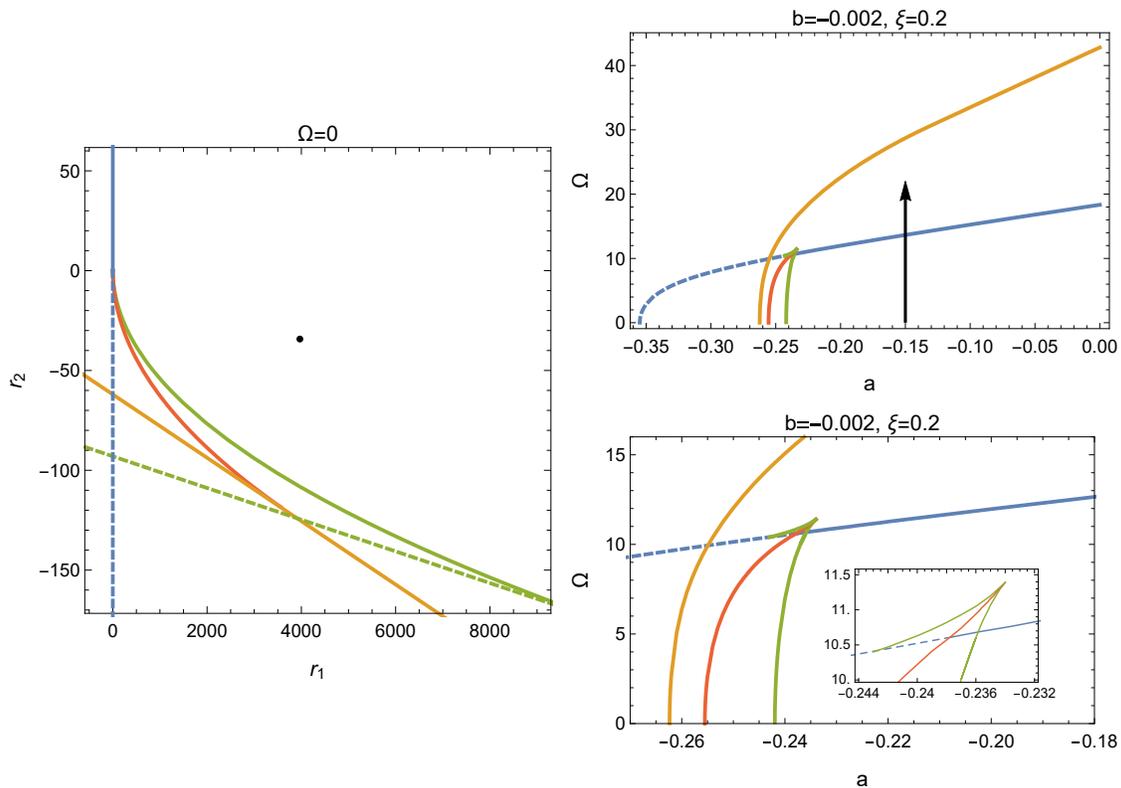}
\end{center}
\caption{System ${}^{12}\mathrm{C}+{}^{12}\mathrm{C}$: In the left we show the parameter space $(r_1,r_2)$ for the case $\Omega=0$. The point corresponds to the mapping of the parameter values of the fitting $a=-0.15$, $b=-0.002$ and $\xi=0.2$. In the upper right plot we show the parameter space $(a,\Omega)$ for $b=-0.002$ and $\xi=0.2$, with the trajectory shown as an vertical arrow at $a=-0.15$. In the bottom right we show a zoom of the bifurcation and Maxwell sets in the $(a,\Omega)$ parameter space. The second order transition separatrix is depicted as a blue line, the bifurcation set as a green line, the Maxwell set as a red line and the stability separatrix as a yellow line. The number of total oscillation quanta, considered to make these plots, is $N+n_0=76$.}
\label{fig.4.1.2}
\end{figure}

A review on the first advances on the study of nuclear molecules, since their discovery \cite{bromley1960,almqvist1960} until 1992, can be found in \cite{satpathy1992}. An historical account of the first eighteen years of the experimental and theoretical research on nuclear molecules can be found in \cite{erb1978}. Since then, there has been a lot of theoretical work in order to understand the phenomenon of nuclear molecules. In particular the antisymmetrized molecular dynamics (AMD) framework, which is a microscopic nuclear model, has been used in the systems ${}^{12}\mathrm{C}+{}^{16}\mathrm{O}$ \cite{taniguchi2009,taniguchi2020} and ${}^{16}\mathrm{O}+{}^{16}\mathrm{O}$ \cite{kimura2004,kimura2020} to study the properties of the molecular states. In Refs. \cite{uegaki2014,uegaki2017} resonances in the system ${}^{12}\mathrm{C}+{}^{12}\mathrm{C}$ are studied with a molecular model and the molecular structures for high spin are described. 
In Refs. \cite{abe1980,abe1979,matsuse1978b,kondo1980} the band crossing model (BCM) was used to study nuclear molecules as resonances for the three systems already mentioned. Here the resonance states appear at the point where the elastic molecular band crosses the inelastic molecular bands.

\begin{figure}[ht]
\begin{center}
\includegraphics[scale=0.8]{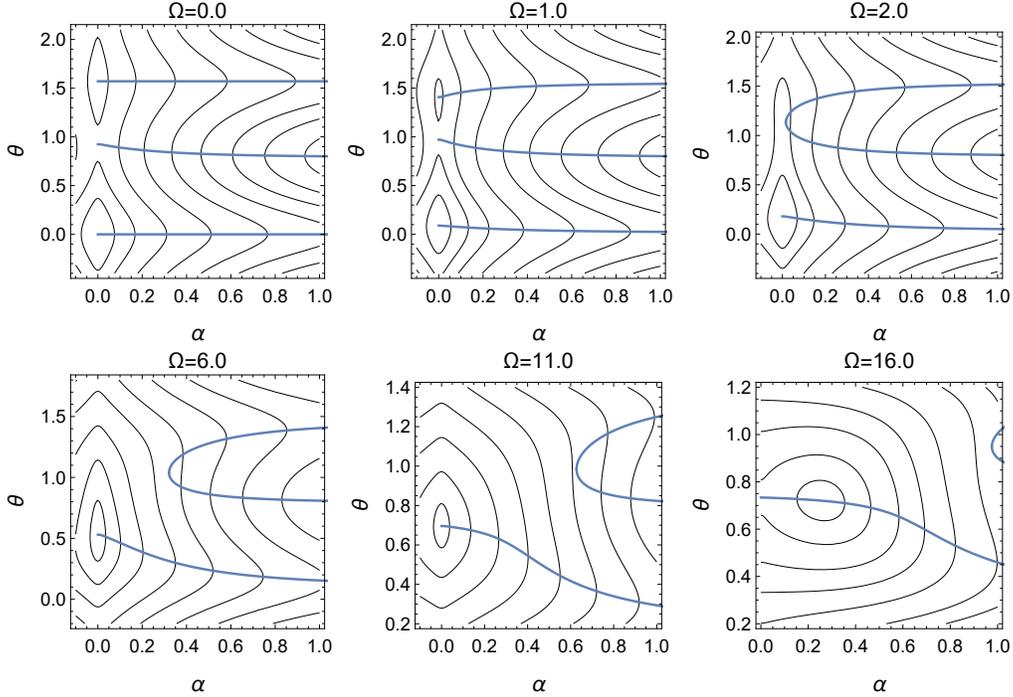}
\end{center}
\caption{System ${}^{12}\mathrm{C}+{}^{12}\mathrm{C}$: Contour plots $(\alpha,\theta)$ of the semiclassical potential for some values of $\Omega$ and $a=-0.15$, $b=-0.002$, and $\xi=0.2$. The blue lines are the critical points of $\theta$ as a function $\alpha$. The number of total oscillation quanta considered to make these plots is $N+n_0=76$.} 
\label{fig.4.1.4}
\end{figure}

\subsection{${}^{12}\mathrm{C}+{}^{12}\mathrm{C}\to {}^{24}\mathrm{Mg}$}
\label{CCMG}
The values that define this cluster system in the SACM are $n_0=12$, $\hbar \omega =12.6\,\mathrm{MeV}$, $(\lambda_1,\mu_1)=(\lambda_2,\mu_2)=(0,4)$, $(\lambda_C,\mu_C)=(4,0)$, $\beta_1=\beta_2=-0.42$ and $\beta_C=0.17$. 
The $(0,4)$ is the irrep of the ${}^{12}\mathrm{C}$ nucleus. The 
cluster irrep is contained as the most compact one
in the product $(0,4) \otimes (0,4)$, which is the less
deformed one. In the combined cluster system
the two clusters remain in their relative orientation,
which results in a $(4,0)$ irrep. Its deformation is used for the
quadrupole-cluster part. In Fig. \ref{fig.4.1.0} we show a schematic illustration of the ${}^{12}\mathrm{C}+{}^{12}\mathrm{C}\to {}^{24}\mathrm{Mg}$ system before and after the QPT.

\begin{figure}[ht]
\begin{center}
\includegraphics[scale=0.8]{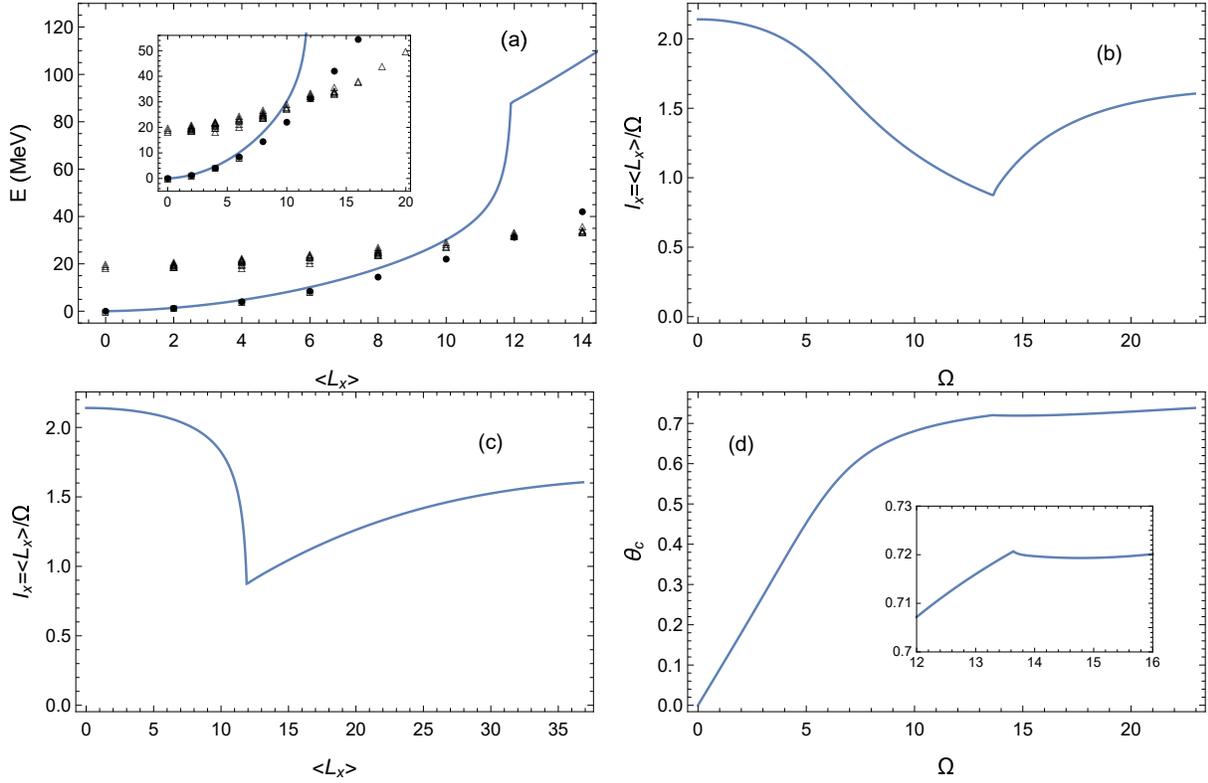}
\end{center}
\caption{System ${}^{12}\mathrm{C}+{}^{12}\mathrm{C}$: 
(a) The energy as a function of $L$ obtained from Eq. (\ref{4.1.1}) and the cranking formalism depicted as a blue line. The theoretical and experimental values \cite{endt1990} of the ground state rotational band are depicted as black circles and blank squares, respectively. The experimental data of resonances associated with molecular states are shown as blank triangles \cite{abbondanno1991}. At about $L\approx 11-12$ the QPT is interpreted as the crossing in energy bands. (b) The moment of inertia as a function of $\Omega$. At the critical value $\Omega_c=13.64$ there is a sharp increase in the moment of inertia signaling the QPT. (c) The moment of inertia as a function of $L$. (d) The critical point $\theta_c$ as a function of $\Omega$. The values used are $a=-0.15$, $b=-0.002$, and $\xi=0.2$. The number of total oscillation quanta considered to make these plots is $N+n_0=76$.} 
\label{fig.4.1.3}
\end{figure}

After performing a fit of the parameters in the Hamiltonian to the experimental spectrum at low energy, we find the following values: $a=-0.15$, $b=-0.002$, $\xi=0.2$, and $t_1=0.7175$. In Table \ref{table12C12C} we show the comparison of the theoretical and experimental energy values of some states.
\begin{table}[ht]
\caption{Energy values comparison for some states of ${}^{24}\mathrm{Mg}$}\label{table12C12C}
\centering
  \begin{tabular}{ c c c c c c }
  \hline
$J^{\pi}$ & $E_{theor} (\mathrm{MeV})$ & $E_{exp} (\mathrm{MeV})$ & $J^{\pi}$ & $E_{theor} (\mathrm{MeV})$ & $E_{exp} (\mathrm{MeV})$  \\ \hline
$0_2^+$ & $7.068$ & $6.432$ & $4_1^+$ & $4.00$ & $4.122$\\ 
$0_3^+$ & $7.171$ & $9.305$ & $4_2^+$ & $6.87$ & $6.010$\\ 
$2_1^+$ & $1.2$ & $1.368$ & $4_3^+$ & $10.567$ & $8.439$\\ 
$2_2^+$ & $4.07$ & $4.238$ & $6_1^+$ & $8.40$ & $8.113$\\ 
$2_3^+$ & $7.767$ & $7.349$ & $6_2^+$ & $11.27$ & $9.528$\\
$2_4^+$ & $8.268$ & $8.654$ & $1_1^-$ & $5.563$ & $7.555$\\ 
$3_1^+$ & $5.27$ & $5.235$ & $1_2^-$ & $9.067$ & $8.438$\\ 
$3_1^-$ & $7.563$ & $7.616$ & $1_3^-$ & $9.731$ & $9.146$\\ \hline
  \end{tabular}
\end{table}
We now may substitute these parameters in the two dimensional semiclassical potential (\ref{3.1.1}) for $\Omega=0$, and obtain the semiclassical potential surface of the ${}^{12}\mathrm{C}+{}^{12}\mathrm{C}\to {}^{24}\mathrm{Mg}$ system 
within the SACM. In Fig. \ref{fig.4.1.4} we show contour plots $(\alpha,\theta)$ of the semiclassical potential. We notice that for $\Omega=0$ the potential has two minima at $(\alpha_c=0,\theta_c=0)$ and $(\alpha_c=0,\theta_c=\pi/2)$, with the dominant minimum located at $(\alpha_c=0,\theta_c=0)$; the third critical point of $\theta$ in (\ref{3.1.4}) corresponds to a saddle point at $\alpha_c=0$. This means that the potential 
presents an absolute minimum when the intercluster axis is aligned to the $z$-axis. Therefore, as the cranking frequency is increased we will track down the evolution of this minimum. In the left side of Fig. \ref{fig.4.1.2} we show the $(r_1,r_2)$ parameter space for $\Omega=0$ with the corresponding mapped point for the values $\{a,b,\xi\}$ of the fit, which falls in the region where the potential is stable and has one spherical minimum at $\alpha=0$. In the right side of Fig. \ref{fig.4.1.2} we plot the parameter space $(a,\Omega)$ for the corresponding values of $b$ 
and $\xi$, and the trajectory of the second order QPT is taken by fixing $a$ 
and increasing $\Omega$ (indicated by the arrow) from zero until the second order transition separatrix is crossed at the critical value $\Omega_c=13.64$. 
At each point in this trajectory, i.e. for each step of $\Omega$, the semiclassical potential (\ref{2.5}) is minimized in the variables $(\alpha,\theta)$, and we obtain a set of critical points $(\alpha_c,\theta_c,\phi_c=0)$ for each 
value of $\Omega$. These values are substituted in (\ref{2.3}) and we can obtain the corresponding values of $L$ as a function of $\Omega$. Similarly, using these results and the relation $L=I\Omega$ we can obtain the moment of inertia as a function of $\Omega$. Before the critical value $\Omega_c$ of the phase transition, the critical point $\alpha_c$ is always zero, and using the limit value in (\ref{2.4}) we can write $\langle {\bd L}_x \rangle=n_0 \sin 2\theta_c$, and the angular critical point $\theta_c$, i.e. the orientation between the clusters, determines the value of the angular momentum for each value of $\Omega$ before the phase transition. In the bottom right plot of Fig. \ref{fig.4.1.3} we show the critical point $\theta_c$ as a function of $\Omega$, and we can see that it grows from zero until it settles near the value of $\pi/4$. We can also see that at the critical value $\Omega_c$ there is a discontinuity in $\theta_c$ signaling the second order phase transition. With these results the value of $L$ where the phase transition takes place can be predicted as the value of $n_0$. After the phase transition the critical point $\alpha_c$ is no longer zero, thus producing a sharp increase in the value of $L$ as $\Omega$ increases. The decrease of the moment of inertia before the phase transition, as seen in the upper right plot of Fig. \ref{fig.4.1.3} can be explained as a change in the orientation of the clusters in their critical position with respect to the $z$-axis, even though the minimal distance between the clusters remains unchanged.
That the moment of inertia is diminishing is an unphysical
development, because for a nuclear molecule, in the classical
sense, the moment of inertia has to increase. The reason for 
the behavior encountered 
is that the cranking axis is still the same,
perpendicular to the old z-axis. By inspecting Fig. \ref{fig.4.1.0}, we note that the 
correct 
rotation axis should be perpendicular to
the axis defined by the angle $\theta$ and connecting 
the centers of both clusters, passing in between the two
clusters which are barely 
touching, which has the largest moment of inertia. 
Thus, the states
obtained within the cranking formalism correspond 
to high lying states and a rotation around an unnatural
axis.

Using a simple rotor model,
the energy of the rotational band is given by
\begin{equation}\label{4.1.1}
E = \frac{\hbar^2}{2I}L(L+1)
~~~,
\end{equation}
and using the previously discussed results and approximations it can be written as: $E=\Omega(\langle {\bd L}_x \rangle +1)/2$. It is possible then to plot the energy as a function of $L$, obtained by using the coherent states and the cranking formalism. In Fig. \ref{fig.4.1.3} we plot the energies
of the rotor model as a function of $L$ alongside with the theoretical values obtained from the fitting as black dots and the experimental data depicted as blank squares \cite{endt1990}; the experimental data of resonances associated with molecular states are shown as blank triangles \cite{abbondanno1991}. For small angular momentum the energy obtained by the cranking formalism matches well the evolution of the states. Then, right before the phase transition there is a sharp increase in energy. This is because for a wide range of values of $\Omega$ the angular critical point is $\theta_c\approx \pi/4$. At the point of the phase transition there is a crossing of energy bands, which matches the place where the experimental data of molecular states cross the ground state rotational band. In the bottom left plot of Fig. \ref{fig.4.1.3} we show the moment of inertia as a function of $L$, and see that it remains almost constant, until about $L\approx 12$ where there is a sharp decrease, followed by a sharp increase after the phase transition. One has to keep in
mind that this is a very simple approach and
that the absolute values of the energies, especially at large angular momentum, may not be reproduced. Nevertheless the change of structure and the transition to a nuclear molecule should be described well.

In Fig. \ref{fig.4.1.4} we show 
contour plots $(\alpha,\theta)$ of the semiclassical potential for different values of $\Omega$. The critical values of $\theta$ as a function of $\alpha$ are depicted as blue lines. We can see that at some value of $\Omega$ the second minimum disappears and after the value $\Omega_c$ the ``spherical'' minimum at $\alpha_c=0$ disappears and becomes a 
``deformed'' minimum at $\alpha_c>0$. (With respect to the
interpretation of ``spherical'' and ``deformed'' see the
discussion above.) The interpretation of this
result is as follows: For low spin the $\alpha$ value is zero
and the united nucleus stays in its real ground state,
which corresponds to the compact irrep obtained, with
the distance of the two clusters small. At
one point, a phase transition appears and the new potential
minimum is at a value of $\alpha > 0$. In this case, the
distance of the two clusters increase abruptly and the
two clusters become well separated, representing a nuclear molecule. 
In the present study, the transition is about $L=12$, which corresponds to the point where the molecular band crosses the ground state rotational band at an energy about $31.2\;\mathrm{MeV}$ as shown in Fig. \ref{fig.4.1.3}.

\begin{figure}[ht]
\begin{center}
\includegraphics[scale=0.8]{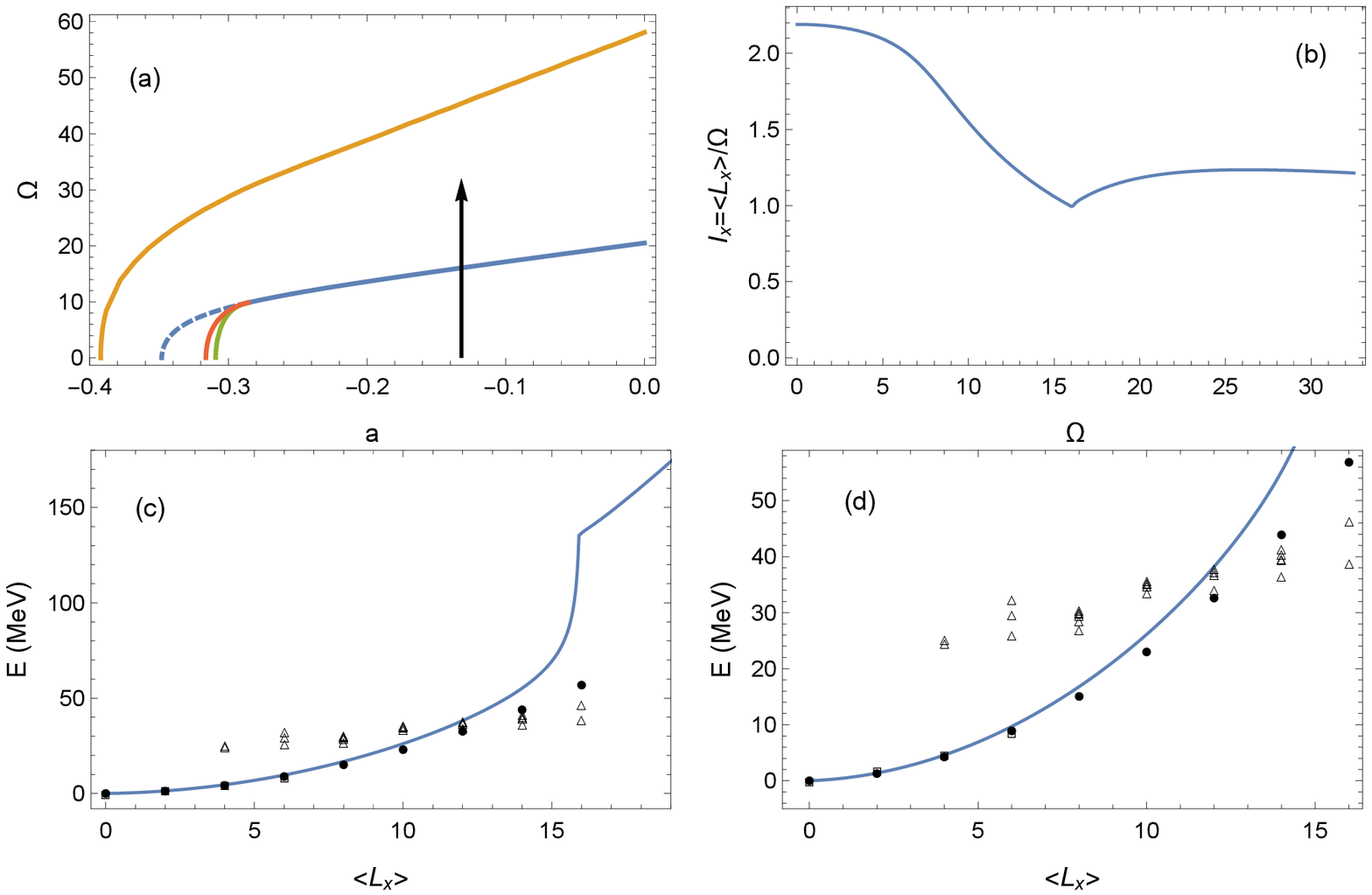}
\end{center}
\caption{
System ${}^{12}\mathrm{C}+{}^{16}\mathrm{O}$: (a) The parameter space $(a,\Omega)$ for the values $b=-0.004$, $\xi=0.209$, and $N+n_0=80$. The black arrow indicates the trajectory taken by increasing the value of $\Omega$ at $a=-0.132$. (b) The moment of inertia as a function of $\Omega$
is plotted, where a sudden increase can be seen at the critical value $\Omega_c=16.06$. 
(c) The energy as a function of $L$ obtained from Eq. (\ref{4.1.1}) and the cranking formalism depicted as a blue line. The theoretical and experimental values \cite{endt1990} of the ground state rotational band are depicted as black circles and blank squares, respectively. The experimental data of resonances associated with molecular states are shown as blank triangles \cite{abbondanno1991}. (d) Zoom of the energy as a function $L$ near the value where the experimental data of molecular states crosses the ground state rotational band. The values of the parameters used in (b), (c) and (d) are $a=-0.132$, $b=-0.004$, and $\xi=0.209$.}
\label{fig.4.2.1}
\end{figure}

An important thing to point out is that for the bifurcation set and Maxwell set to appear in the $(a,\Omega)$ parameter space that value of $N$ must be large. In Fig. \ref{fig.4.1.2} we considered $N=64$. However, 
{\it the critical value} $\Omega_c$ 
{\it does not depend on $N$}, and it appears that no matter how large $N$ is, the bifurcation and Maxwell set will not be crossed for the values of the parameters $(a,b,\xi)$ that result in the fitting of the Hamiltonian to experimental data. Nonetheless it is important to consider that these separatrices exist for large values of $N$ and that they may come into play for some other cluster systems.

Once the geometrical molecular structure is obtained
(relative orientation and distance of the clusters), we can
proceed in using the geometrical model of a nuclear molecule
\cite{NM1,NM2} and determine its spectrum, which we intend to do in a future publication.

\subsection{${}^{12}\mathrm{C}+{}^{16}\mathrm{O}\to {}^{28}\mathrm{Si}$}
\label{COSi}
The values that define this cluster system in the SACM are $n_0=16$, $\hbar \omega =12.11\,\mathrm{MeV}$, $(\lambda_1,\mu_1)=(0,4)$, $(\lambda_2,\mu_2)=(0,0)$, $(\lambda_C,\mu_C)=(0,4)$, $\beta_1=-0.42$, $\beta_2=0.0$, and $\beta_C=-0.14$. For this system the values of the parameters obtained after the fit are: $a=-0.132$, $b=-0.004$ and $\xi=0.209$, $t_1=0.0$. In Table \ref{table12C16O} we show the comparison of the theoretical and experimental values of some states.
\begin{table}[ht]
\caption{Energy values comparison for some states of ${}^{28}\mathrm{Si}$}\label{table12C16O}
\centering
  \begin{tabular}{ c  c  c  c  c  c }
  \hline
$J^{\pi}$ & $E_{theor} (\mathrm{MeV})$ & $E_{exp} (\mathrm{MeV})$ & $J^{\pi}$ & $E_{theor} (\mathrm{MeV})$ & $E_{exp} (\mathrm{MeV})$  \\ \hline
$0_2^+$ & $5.0717$ & $4.9799$ & $6_1^+$ & $8.778$ & $8.5435$\\ 
$0_3^+$ & $6.3031$ & $6.6907$ & $1_1^-$ & $3.5179$ & $8.9048$\\ 
$2_1^+$ & $1.254$ & $1.779$ & $3_1^-$ & $5.6079$ & $6.8787$\\ \
$2_2^+$ & $6.3257$ & $7.3805$ & $4_1^-$ & $7.2799$ & $8.4133$\\
$2_3^+$ & $6.3257$ & $7.4162$ & $1_1^+$ & $12.5578$ & $8.3283$\\ 
$4_1^+$ & $4.18$ & $4.6178$ & $3_1^+$ & $7.5797$ & $6.2762$\\ 
$4_2^+$ & $9.2517$ & $6.8876$ & $3_2^+$ & $8.8111$ & $7.799$\\ \hline
  \end{tabular}
\end{table}
Substituting these values of the parameters in the potential (\ref{3}) for some values of $\Omega$ yields a potential surfaces and in the case $\Omega=0$ we see that the dominant minimum is located at $(\alpha_c=0,\theta_c=\pi/2)$. The values of the physical observables are obtained in the same way as in the previous subsection. In the top left of Fig. \ref{fig.4.2.1} the $(a,\Omega)$ parameter space 
is shown for the corresponding values of $b$ and $\xi$ and the trajectory taken at $a=-0.132$ as the value of $\Omega$ is increased. In the top right of Fig. \ref{fig.4.2.1} the moment of inertia as a function of $\Omega$ is plotted. We see how for small values of $\Omega$ it stays almost constant and then it starts to decrease until the QPT, where a sharp increase starts to happen at the critical value $\Omega_c=16.06$. In the bottom row we plot the energy of the ground state rotational band as a function of $L$, where the second order QPT can be seen at about $L\approx 15-16$. In this case the crossing of the ground state rotational band with the molecular resonance band occur at an energy about $43.89\,\mathrm{MeV}$ which corresponds to $L=14$.
Similar to Fig. \ref{fig.4.1.3} an unnatural low moment of inertia
appears in the molecular band shown in Fig \ref{fig.4.2.1}, and its origin is explained in the main text.

\begin{figure}[ht]
\begin{center}
\includegraphics[scale=0.8]{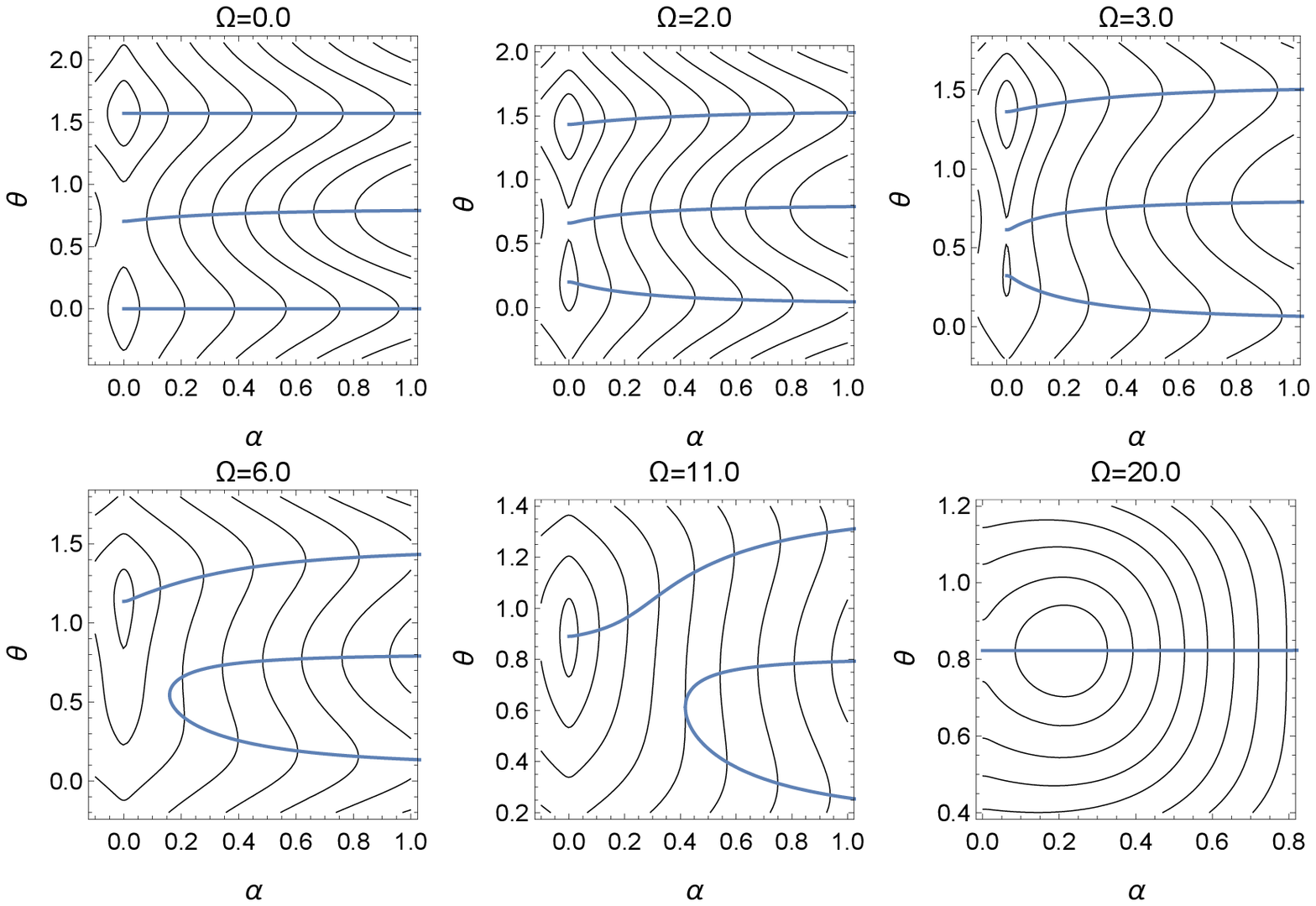}
\end{center}
\caption{System ${}^{12}\mathrm{C}+{}^{16}\mathrm{O}$: Contour plots $(\alpha,\theta)$ of the semiclassical potential for some values of $\Omega$ and $a=-0.132$, $b=-0.004$, and $\xi=0.209$. The blue lines are the critical points of $\theta$ as a function $\alpha$. For $\Omega=0$ the dominant minimum is located $(\alpha_c=0,\theta_c=\pi/2)$. The number of total oscillation quanta considered to make these plots is $N+n_0=80$.} 
\label{fig.4.3.1}
\end{figure}

In Fig. \ref{fig.4.3.1} the contour plots for different values of $\Omega$ of the potential surface
is plotted. The blue line represent the critical values of $\theta$ as a function of $\alpha$, indicating the steepest descend of the surface. The dominant minimum is located at $(\alpha_c=0,\theta_c=\pi/2)$, and as the $\Omega$ frequency increases the other minimum disappears, and then at the critical value $\Omega_c$ the minimum located at $\alpha_c=0$ disappears and becomes a minimum located at $\alpha_c>0$.

\subsection{${}^{16}\mathrm{O}+{}^{16}\mathrm{O}\to {}^{32}\mathrm{S}$}
\label{OOS}
This case is different than the previous two. If we approximate it in a similar fashion we find that the values that define this cluster system in the SACM are $n_0=20$, $\hbar \omega=11.69\,\mathrm{MeV}$, $(\lambda_1,\mu_1)=(\lambda_2,\mu_2)=(\lambda_C,\mu_C)=(0,0)$, and $\beta_1=\beta_2=0.0$. However, if this is the case and we construct the space of the ${}^{32}\mathrm{S}$ nucleus in the SACM, we find that the ground state irrep cannot be generated from the product. In this case the concept of forbiddenness must be taken int account, in which one of the ${}^{16}\mathrm{O}$ clusters is assumed to be in an excited state. Following \cite{tesislic} 
one of the clusters has an irrep $(\lambda_2,\mu_2)=(10,2)$ and $8$ additional quanta of excitation, meaning that the Wildermuth condition of the system is now $n_0=12$.
Additionally the deformation parameter for this cluster is no longer zero and its value is approximated using the relation in (\ref{betac}), which gives us a value about $\beta_2=0.66$. The parameter deformation of the cluster system is calculated in a similar way and results in $\beta_C=0.26$.

The values of the Hamiltonian parameters obtained after the fitting with the experimental data are: $a=-0.107$, $b=-0.0033$, $\xi=0.215$, and $t_1=0.7007$; and in Table \ref{table16O16O} we show the comparison of the theoretical and experimental values of some states.
\begin{table}
\caption{Energy values comparison for some states of ${}^{32}\mathrm{S}$}  \label{table16O16O}
\centering
  \begin{tabular}{ c  c  c  c  c  c }
  \hline
$J^{\pi}$ & $E_{theor} (\mathrm{MeV})$ & $E_{exp} (\mathrm{MeV})$ & $J^{\pi}$ & $E_{theor} (\mathrm{MeV})$ & $E_{exp} (\mathrm{MeV})$  \\ \hline
$0_2^+$ & $1.8284$ & $3.7784$ & $1_1^+$ & $8.8686$ & $4.6953$\\ 
$0_3^+$ & $6.8947$ & $7.5357$ & $1_2^+$ & $9.7128$ & $7.0014$\\ 
$2_1^+$ & $1.29$ & $2.2305$ & $1_1^-$ & $5.5605$ & $5.7968$\\ 
$2_2^+$ & $3.1184$ & $4.2818$ & $2_1^-$ & $8.2871$ & $6.2229$\\ 
$2_3^+$ & $4.0928$ & $5.5482$ & $3_1^-$ & $7.7105$ & $5.0062$\\ 
$4_1^+$ & $4.3$ & $4.4591$ & $4_1^-$ & $11.2971$ & $6.6217$\\ 
$4_2^+$ & $6.1284$ & $6.411$ & $3_1^+$ & $5.3828$ & $5.4126$\\ 
$6_1^+$ & $9.03$ & $8.3464$ &  &  & \\ \hline
  \end{tabular}
\end{table}

\begin{figure}[ht]
\begin{center}
\includegraphics[scale=0.8]{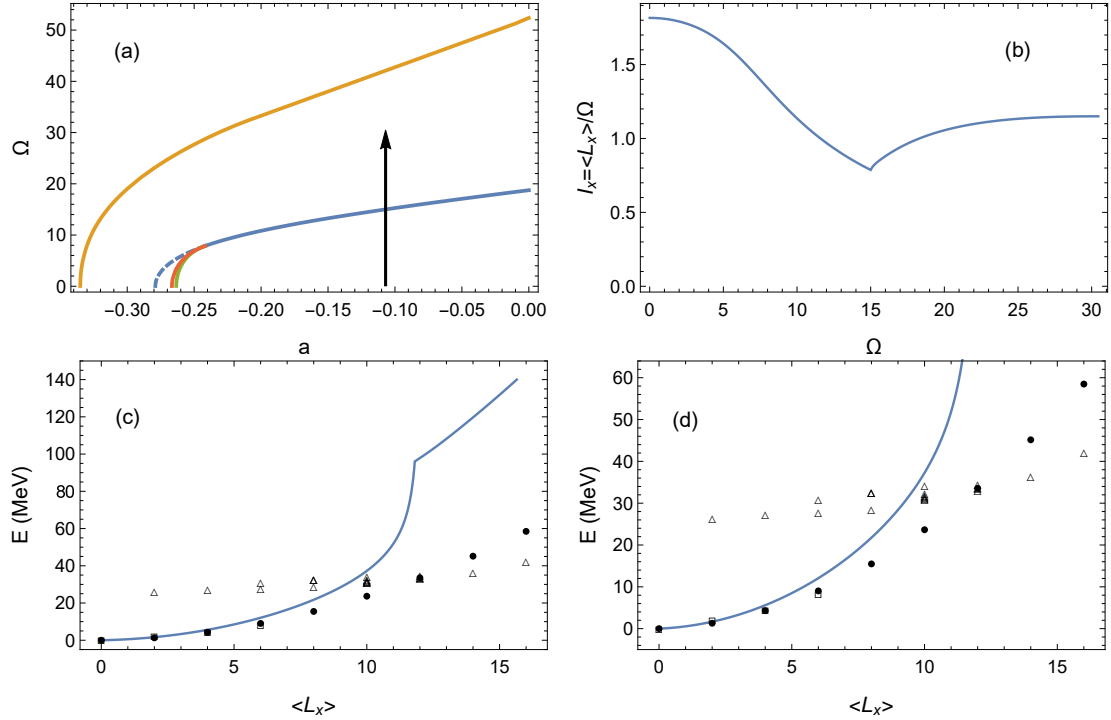}
\end{center}
\caption{
System ${}^{16}\mathrm{O}+{}^{16}\mathrm{O}$: (a) The parameter space $(a,\Omega)$ for the values $b=-0.0033$, $\xi=0.215$, and $N+n_0=76$. The black arrow indicates the trajectory taken by increasing the value of $\Omega$ at $a=-0.107$. (b) The moment of inertia as a function of $\Omega$, where a sudden increase can be seen at the critical value $\Omega_c=15.00$ indicating a second order QPT. 
(c) The energy as a function of $L$ obtained using Eq. (\ref{4.1.1}) and the cranking formalism as a blue line. At about $L\approx 11-12$ the second order QPT takes place and can be seen as the crossing of energy bands. 
The theoretical and experimental values \cite{endt1990} of the ground state rotational band are depicted as black circles and blank squares, respectively. The experimental data of resonances associated with molecular states are shown as blank triangles \cite{abbondanno1991}. (d) Zoom of the energy as a function $L$ near the value where the experimental data of molecular states crosses the ground state rotational band. The values of the parameters used in (b), (c) and (d) are $a=-0.107$, $b=-0.0033$, and $\xi=0.215$.}
\label{fig.4.3.3}
\end{figure}

In the upper left plot of Fig. \ref{fig.4.3.3} we show the $(a,\Omega)$ parameter space and the trajectory taken, and in the upper right we plot the moment of inertia as a function of $\Omega$. At the critical value $\Omega_c=15.00$ a second order QPT occurs as seen by a sharp 
decrease in the moment of inertia.
The reason of this behavior and its interpretation
was already given in the two former systems and it is related
to a wrong rotation axis after the phase transition. 
In the bottom row of Fig. \ref{fig.4.3.3} we plot the energy of the ground state rotational band as a function of $L$, where the second order QPT can be seen at about $L\approx 11-12$. The crossing of the ground state rotational band with the molecular resonance band occur at an energy about $33.54\,\mathrm{MeV}$ which corresponds to $L=12$.

\section{Conclusions}
\label{conclusions}
The cranking mechanism was applied to a two-cluster system of light nuclei with an algebraic cluster model, known as the SACM.
The motivation is to find a unifying geometrical description
of phase transitions in two-cluster systems in excited nuclei.

At low energy the nucleus exhibits a compact structure, but as the angular frequency $\Omega$ of the cranking is increased a point is reached at the critical value $\Omega_c$, where the two clusters are physically separated forming a nuclear molecule. Using a semiclassical analysis with the coherent states of the SACM, this was identified as a second order phase transition, describing the transition from a compact structure to a nuclear molecule. Three systems were investigated: ${}^{12}\mathrm{C}+{}^{12}\mathrm{C}$, ${}^{12}\mathrm{C}+{}^{16}\mathrm{O}$, and ${}^{16}\mathrm{O}+{}^{16}\mathrm{O}$. The value of the angular momentum at which the phase transition takes place is in accordance to the crossing of the ground state rotational band and the experimental values of the molecular resonances.
Though, the onset of the phase transition is well reproduced,
the excited states after the phase transition are not.
They correspond to high lying states with an unnatural
rotation 
axis, with a too low moment 
of inertia. 
The correct rotation axis is tilted such that
it is orthogonal to the axis indicated by the angle $\theta$
(corresponding to an axis tilted by this angle with respect to
the old z-axis and which is connecting the centres
of both clusters)
and passing between the unchanged two clusters, 
which are just touching.

It was possible to construct separatrices in the parameter space $(a,\Omega)$ for each of the systems. Two separatrices are of particular interest in the study of phase transitions: The Maxwell set and the second order transition separatrix, corresponding to first and second order QPTs, respectively. In the systems studied here the fitted parameters of the Hamiltonian mapped to a point near the second order transition separatrix and it was crossed at some critical value $\Omega_c$ as the cranking parameter was increased. It would be interesting if for some other system the fitted parameters mapped to a point near the Maxwell set for a first order QPT to take place.

In the future it would be interesting also to determine the
spectrum of the nuclear molecular systems encountered
using the model published in \cite{NM1,NM2}. We also intend to study heavy nuclei using the recent SACM extension to this range \cite{chavez2021}. We plan to study the position of phase transitions for different clusterizations in ${}^{236}\mathrm{U}$ and hope to contribute to a better understanding of the fission process and of the SACM for heavy nuclei.
Studies of preferences 
on nuclear clusterization, using the SACM and 
a specific definition
of the forbiddenness, can be found in 
\cite{csehpref1,csehpref2,csehpref3}.

\section*{Acknowledgment}

P.O.H. acknowledges financial help from DGAPA (IN100421). D.S.L.R. acknowledges financial support from a scholarship (No. 728381) received from CONACyT.


\vskip0.5cm

\begin{thebibliography}{99}


\bibitem{cejnar2010} P. Cejnar, J. Jolie, R.F. Casten, Rev. Mod. Phys. 82 (2010) 2155.

\bibitem{lopezmoreno1996b} E. L\'opez-Moreno, O. Casta\~nos, Rev. Mex. Fis. S 42 (1996) 163.

\bibitem{lopezmoreno1998} E. L\'opez-Moreno, O. Casta\~nos, Rev. Mex. Fis. S 44 (1998) 48.

\bibitem{ring} P. Ring, P. Schuck, The
Nuclear Many-Body Problem, Springer, Heidelberg, 1960.

\bibitem{cseh1993} J. Cseh, G. Levai, W. Scheid, Phys. Rev. C 48 (1993) 1724.

\bibitem{chavez2021} P.O. Hess, L.J. Ch\'avez-Nu\~nez, E. Phys. J. A 57 (2021) 146.

\bibitem{NM1} P.O. Hess, W. Greiner, W.T. Pinkston,
Phys. Rev. Lett. 53 (1984) 1535.

\bibitem{NM2} P.O. Hess, W. Greiner, Il Nuovo
Cimento 83 (1984) 76.

\bibitem{hess1996} P.O. Hess, G. L\'evai, J. Cseh, Phys. Rev. C 54 (1996) 2345.

\bibitem{lohrrobles2019} D.S. Lohr-Robles, E. L\'opez-Moreno, P.O. Hess, Nucl. Phys. A 992 (2019) 121629.

\bibitem{lohrrobles2021} D.S. Lohr-Robles, E. L\'opez-Moreno, P.O. Hess, Nucl. Phys. A 1016 (2021) 122335.

\bibitem{cindro1981} N. Cindro, Riv. Nuovo Cim. 4 (1981) 1.  

\bibitem{abbondanno1991} U. Abbondanno, Report No. INFN/BE-91/11, Trieste (1991).

\bibitem{cseh1992} J. Cseh, Phys. Lett. B 281 (1992) 1737.

\bibitem{cseh1994} J. Cseh, G. L\'evai, Ann. Phys. 230 (1994) 165.

\bibitem{elliott1958a} J.P. Elliott, Proc. Roy. Soc. A 245 (1958) 128.

\bibitem{elliott1958b} J.P. Elliott, Proc. Roy. Soc. A 245 (1958) 562.

\bibitem{vibron1} F. Iachello, Phys. Rev. C 23 (1981) 2778.

\bibitem{vibron2} F. Iachello, Chem. Phys. Lett. 78 (1981) 581.

\bibitem{vibron3} F. Iachello, R.D. Levine, J. Chem. Phys. 77 (1982) 3046.

\bibitem{blomqvist1968} J. Blomqvist, A. Molinari, Nucl. Phys. A 106 (1968) 545.

\bibitem{wildermuth} K. Wildermuth, Y.C. Tang, A Unified Theory of the Nucleus, Friedr. Vieweg \& Sohn Verlagsgesselschaft mbH, Braunschweig, 1977.

\bibitem{moraleshernandez2012} G.E. Morales-Hern\'andez, H. Y\'epez-Mart\'inez, P.O. Hess, J. Phys.: Conf. Ser. 387 (2012) 012019.

\bibitem{lopezmoreno2016} E. L\'opez-Moreno, G.E. Morales-Hern\'andez, P.O. Hess, H. Y\'epez-Mart\'inez, J. Phys.: Conf. Ser. 730 (2016) 012017.

\bibitem{morales2012} G.E. Morales Hern\'andez, Bachelor
thesis UNAM (2012).

\bibitem{morales2015} G.E. Morales Hern\'andez, Master thesis UNAM (2015).

\bibitem{yepezmartinez2012a} H. Y\'epez-Mart\'inez, P.R. Fraser, P.O. Hess, G. L\'evai, Phys. Rev. C 85 (2012) 014316.

\bibitem{rowe} D. J. Rowe, Rep. Prog. Phys. 48 (1985), 1419.

\bibitem{castanos1988} O. Casta\~nos, J.P. Draayer, Y. Leschber, Z. Phys. A 329 (1988) 33.

\bibitem{gilmore} R. Gilmore, Catastrophe Theory for Scientists and Engineers, Wiley, New York, 1981.

\bibitem{cejnar2002} P. Cejnar, Phys. Rev. C 65 (2002) 044312.

\bibitem{cejnar2003} P. Cejnar, Phys. Rev. Lett. 90 (2003) 112501.

\bibitem{cejnar2004} P. Cejnar, J. Jolie, Phys. Rev. C 69 (2004) 011301. 

\bibitem{schaaser1986} H. Schaaser, D.M. Brink, Nucl. Phys. A 452 (1986) 1.

\bibitem{yepezmartinez2008} H. Y\'epez-Mart\'inez, P.O. Hess, Rev. Mex. Fis. S 54 (2008) 69.

\bibitem{smirnov1984} Yu.F. Smirnov, Yu.M. Tchuvil'sky, Phys. Lett. B 134 (1984) 25.

\bibitem{yepezmartinez2015} H. Y\'epez-Mart\'inez, P.O. Hess, J. Phys. G: Nucl. Part. Phys. 42 (2015) 095109.

\bibitem{bromley1960} D.A. Bromley, J.A. Kuehner, E. Almqvist, Phys. Rev. Lett. 4 (1960) 365

\bibitem{almqvist1960} E. Almqvist, D.A. Bromley, J.A. Kuehner, Phys. Rev. Lett. 4 (1960) 515.

\bibitem{satpathy1992} L. Satpathy, Prog. Part. Nucl. Phys. 29 (1992) 327.

\bibitem{erb1978} K.A. Erb, D.A. Bromley, J. Weneser, Comments Nucl. Part. Phys. 8 (1978) 111.

%
%
%

\bibitem{taniguchi2009} Y. Taniguchi, Y. Kanada-En'yo, M. Kimura, Phys. Rev. C 80 (2009) 044316.

\bibitem{taniguchi2020} Y. Taniguchi, M. Kimura, Phys. Lett. B 800 (2020) 135086.

\bibitem{kimura2004} M. Kimura, H. Horiuchi, Phys. Rev. C 69 (2004) 051304(R).

\bibitem{kimura2020} M. Kimura, Y. Taniguchi, Phys. Rev. C 102 (2020) 024325.

\bibitem{uegaki2014} E. Uegaki, Y. Abe, J. Phys.: Conf. Ser. 569 (2014) 012091.

\bibitem{uegaki2017} E. Uegaki, Y. Abe, J. Phys.: Conf. Ser. 863 (2017) 012048.

\bibitem{abe1980} Y. Abe, Y. Kondo, T. Matsuse, Prog. Theor. Phys. Suppl. 68 (1980) 303.

\bibitem{abe1979} Y. Abe, T. Matsuse, Y. Kondo, Phys. Rev. C 19 (1979) 1365.

\bibitem{matsuse1978b} T. Matsuse, Y. Abe, Y. Kondo, Prog. Theor. Phys. 59 (1978) 1904.

\bibitem{kondo1980} Y. Kondo, D.A. Bromley, Y. Abe, Phys. Rev. C 22 (1980) 1068.

\bibitem{endt1990} P.M. Endt, Nucl. Phys. A 521 (1990) 1.


%

\bibitem{tesislic} L.J. Ch\'avez-Nu\~nez, Bachelor's thesis UNAM, (2019).

\bibitem{csehpref1} A. Algora, J. Cseh, P.O. Hess, J. Phys. G: Nucl. Part. Phys. 24 (1998) 2111.

\bibitem{csehpref2} A. Algora, J. Cseh, P.O. Hess, J. Phys. G: Nucl. Part. Phys. 25 (1999) 775.

\bibitem{csehpref3} A. Algora, J. Cseh, P.O. Hess, M. Hunyadi, Heavy Ion Phys. 13 (2001) 145.

\end{thebibliography}
\end{document}